\documentclass[preprint]{aastex}        
\usepackage{graphicx,times,natbib,lscape,longtable}

\newcommand{\sersic}{{S\'{e}rsic }}

\def\arcsec{$^{\prime\prime}$}

\def\spose#1{\hbox to 0pt{#1\hss}}
\def\lta{\mathrel{\spose{\lower 3pt\hbox{$\mathchar"218$}}
     \raise 2.0pt\hbox{$\mathchar"13C$}}}

\shorttitle{Clumps in Star-Forming Galaxies at z$\sim$2}
\shortauthors{Guo et al.}

\begin{document}

\title{Multi-Wavelength View of Kiloparsec-Scale Clumps in Star-Forming Galaxies at z$\sim$2}
\author{Yicheng Guo$^{1,2}$, Mauro Giavalisco$^{1}$, Henry C. Ferguson$^{3}$,
Paolo Cassata$^{1,4}$, Anton M. Koekemoer$^{3}$}
\affil{$^1$ Astronomy Department, University of Massachusetts,
710 N. Pleasant St., Amherst, MA 01003, USA}
\affil{$^2$ email: {\texttt yicheng@astro.umass.edu}}
\affil{$^3$ Space Telescope Science Institute, 3700 San Martin Drive,
Baltimore, MD, 21218, USA}
\affil{$^4$ Laboratoire d'Astrophysique de Marseille, 38 avenue F. Joliot-Curie,
F-13388 Marseille cedex 13, France}



\begin{abstract} 

This paper studies the properties of kiloparsec-scale clumps in star-forming
galaxies (SFGs) at z$\sim$2 through multi-wavelength broad band photometry. A
sample of 40 clumps is identified from HST/ACS z-band images through
auto-detection and visual inspection from 10 galaxies with 1.5$<$z$<2.5$ in the
Hubble Ultra Deep Field (HUDF), where deep and high-resolution HST/WFC3 and ACS
images enable us to resolve structures of z$\sim$2 galaxies down to kiloparsec
(kpc) scale in the rest-frame UV and optical bands and to detect clumps toward
the faint end.
The physical properties of clumps are measured through fitting spatially
resolved seven-band (BVizYJH) spectral energy distribution (SED) to models.  On
average, the clumps are blue and have similar median rest-frame UV--optical
color as the diffuse components of their host galaxies, but the clumps have
large scatter in their colors. Although the SFR--stellar mass relation of
galaxies is dominated by the diffuse components, clumps emerge as regions with
enhanced specific star formation rates (SSFRs), contributing individually
$\sim$10\% and together $\sim$50\% of the star formation rate (SFR) of the host
galaxies. However, the contributions of clumps to the rest-frame UV/optical
luminosity and stellar mass are smaller, typically a few percent individually
and $\sim$20\% together. On average, clumps are younger by 0.2 dex and denser
by a factor of eight than diffuse components. Clump properties have obvious radial
variations in the sense that central clumps are redder, older, more extincted,
denser, and less active on forming stars than outskirt clumps. 
Our results are broadly consistent with a widely held view that clumps are
formed through gravitational instability in gas-rich turbulent disks and would
eventually migrate toward galactic centers and coalesce into bulges. Roughly
40\% of the galaxies in our sample contain a massive clump that could be
identified as a proto-bulge, which seems qualitatively consistent with such a
bulge-formation scenario.

\end{abstract}

\section{Introduction}
\label{intro}

Understanding when and how the Hubble sequence observed today was formed
remains an unsolved problem in astronomy. To answer it requires tracking the
evolution of galaxy morphology and kinematics back to high redshift. Thanks to
the emergence of facilities with deep sensitivity and high resolution, e.g.,
HST/ACS, NICMOS and WFC3, galaxy morphology and structure now can be resolved
into kpc scale, allowing a study on the properties of sub-structures at
redshift of 2 or higher
\citep[e.g.,][]{elmegreen05,elmegreen07,elmegreen09a,elmegreen09b,gargiulo11,ycguo11peg,szomoru11}.
An interesting population of galaxies at z$\sim$2, the peak of cosmic star
formation activity, is star-forming galaxies (SFGs) that have star formation
rate (SFR) of tens to a few hundred ${\rm M_\odot yr^{-1}}$. This population
contributes a large fraction of the cosmic SFR and stellar mass density at the
epoch \citep[e.g.,][]{daddi07a,grazian07,reddy08,ly11,rodighiero11,ycguo12vjl}. In contrast to
passively-evolving galaxies at z$\sim$2, which have spheroid-like and compact
morphology
\citep[e.g.,][]{daddi05,trujillo06,trujillo07,vandokkum08,vandokkum10,cassata10},
SFGs exhibit a wide diversity on their morphology, from smooth disk-like
structure to irregular or merger-like structure
\citep[e.g.,][]{conselice04,conselice08,lotz06,ravindranath06}. A common and
unique feature of SFGs at z$\sim$2 is the existence of giant kpc-scale clumps,
which are associated with all types of SFGs at z$\sim$2
\citep[e.g.,][]{elmegreen05,elmegreen07,elmegreen09a,bournaud08,genzel08,genzel11,fs11b}
but resemble neither the starburst regions nor bulges of low redshift SFGs.

These giant clumps are mostly identified in high-resolution HST optical ACS
images \citep[e.g.,][]{elmegreen05,elmegreen07}, which probe the rest-frame UV
region at z$>$1 at physical spatial resolution of $\sim$1 kpc. They are also
detected in HST near-infrared (NIR) images of NICMOS
\citep{elmegreen09a,fs11b}, which observe the rest-frame optical wavelengths at
z$>$1. Rest-frame optical line emissions from NIR integral field spectroscopy
observations of z$\sim$2 disk galaxies also reveal such giant clumps
\citep{genzel08,genzel11}. Clumps are studied at even higher effective spatial
resolution, up to a few $\sim$100 pc, in strongly lensed objects at 1$<$z$<$3
through rest-optical or CO line emission \citep[e.g.,][]{jones10,swinbank10}.
However, the fraction of galaxies at z$\gtrsim$2 that show clumpy structures is
still uncertain.  \citet{lotz06,ravindranath06} and Ravindranath et al. (2011,
in prep.) concluded that clumpy galaxies are only about 30\% of the population
at z$\sim$3, while \citet{elmegreen07} argued that the dominant morphology for
z$\gtrsim$2 is clumpy galaxies. \citet{wuyts12} measured the fraction of
clumpy galaxies in a mass-complete sample of SFGs at z$\sim$2 by using
multi-waveband images and stellar mass maps. They found that the fraction
depends sensitively on the light/mass map used to identify the clumps,
decreasing from about 75\% for clumps selected through rest-frame 2800 \AA
images to about 40\% for clumps selected through rest-frame V-band images or
stellar mass maps.

The formation and subsequent evolution of these giant clumps, thought to be
linked with the formation of bulges and thick disks in today's massive
galaxies, are still unknown. In a widely held hypothesis based on theoretical
works and numerical simulations, these kpc-scale clumps are formed through
gravitational instability in gas-rich turbulent disks
\citep[e.g.,][]{noguchi99,immeli04a,immeli04b,elmegreen08,dekel09,ceverino10,ceverino12}.
This scenario is supported by the fact that high-redshift galaxies are
gas-rich, with the gas--to--baryonic fraction of 20\% to 80\%
\citep[e.g.,][]{erb06,genzel08,tacconi08,tacconi10,fs09,daddi10}, possibly as a
result of smooth and continuous accretion of cold gas flow
\citep{keres05,rauch08,dekel09gas,cresci10,steidel10,giavalisco11}. Due to the
clump interactions and dynamical friction against the surrounding disks, these
clumps would migrate toward the gravitational centers of their host galaxies
and eventually coalesce into a young bulge as the progenitor of today's bulges
or be disrupted by either tidal force or stellar feedback to form part of a
thick disk \citep[e.g.,][]{bournaud09,dekel09,murray10,genel12}.

To confirm or set constrains on the above scenarios, one needs accurate
measurements on the physical properties of clumps, such as stellar mass, age,
SFR, and star formation history (SFH) as well as their radial variations across
the host galaxies. Such knowledge, unfortunately, is lacking and difficult to
obtain, because it requires spatially resolved multi-band photometry or
spatially resolved spectroscopy of each clump. In the work of \citet{fs11b},
only two of their six galaxies have marginally enough information to enable a
rough measurement on M/L and age for clumps, one with a single ACS/F814W -
NIC2/F160W color and the other with H$\alpha$ equivalent width measured by the
adaptive optical assisted SINFONI. Their finding of evidence of a systematic
trend of older clumps at smaller galactocentric radius is consistent with the
above scenarios, but suffers from a small number statistics.
\citet{elmegreen09a} studied clump clusters and chain galaxies in the HUDF with
{\it HST}/ACS and NICMOS images. They fit stellar population models to the
multi-band photometry of clumps and bulges to derive their stellar masses and
ages. However, the resolution of their NICMOS images is three times poorer than
that of the ACS images so that obtaining uniform multi-band photometry for
individual clump is affected by the blend of clumps in the low-resolution
images. Recently, \citet{wuyts12} performed a detailed analysis on the
spatially resolved colors and stellar populations of a large and complete
sample of massive SFGs at $0.5<z<2.5$ in the ERS and CANDELS-Deep region of
GOODS-South. The multi-wavelength images of {\it HST}/ACS and WFC3 allow them
to study light maps in both rest-frame UV and rest-frame optical with similar
spatial resolutions. Their results are consistent with an inside-out disk
growth scenario with a ``Christmas tree'' model in which giant star-forming
clumps are formed through gravitational instabilities in gas-rich disks, and
then are quickly disrupted through feedbacks. Alternatively, the presence of
blue and young star-forming clumps superposed on a redder underlying disk in
their SFGs is also consistent with the inward migration scenario. However,
\citet{wuyts12} focused on regions with excess surface brightness and did not
subtract the diffuse light of host galaxies from clumps.

In this paper, we try to measure the physical properties of clumps in SFGs at
z$\sim$2 in a sufficiently large sample, exploiting the advantage of high
resolution and deep sensitivity of HST/ACS and WFC3 images in the HUDF. The
recently available WFC3 images allow us to measure spatially resolved
multi-band photometry across the rest-frame UV and optical regions at this
redshift. Especially, the WFC3 F160W image observes light with rest-frame
wavelengths longer than the Balmer break, enabling an accurate measurement on
the age and stellar mass of clumps. On the other hand, the high resolution of
WFC3 images (0.15\arcsec) enables us to resolve into $\sim$1 kpc scale, a
typical size of clumps, at z$\sim$2 to effectively separate the light from
clumps and their surrounding disks. In this work, we compare the properties of
clumps to those of their surrounding materials (Sec. \ref{prp}) and also study
the radial variations of clump properties across their host galaxies (Sec.
\ref{prpvar}). 

Throughout the paper, we adopt a flat ${\rm \Lambda CDM}$ cosmology with
$\Omega_m=0.3$, $\Omega_{\Lambda}=0.7$ and use the Hubble constant in terms of
$h\equiv H_0/100 {\rm km~s^{-1}~Mpc^{-1}} = 0.70$.  All magnitudes in the paper
are in AB scale \citep{oke74} unless otherwise noted.

\section{The Data and Sample Selection}
\label{data}

\subsection{Images and Catalogs}
\label{data:image}

The ultra-deep ACS images in the HUDF \citep{beckwith06hudf} cover an area
roughly equal to the footprint of the ACS/WFC FOV in the same four filters as
the GOODS ACS program, namely F435W (B), F606W (V), F775W (i), and F850LP (z)
down to a depth of 29.4, 29.8, 29.7, and 29.0 mag (5$\sigma$,
0.35\arcsec-diameter aperture), respectively. We use the publicly available
images, which have been re-binned to the same pixel scale as the GOODS/ACS
mosaic, namely 0.03\arcsec\ /pixel ($0.6 \times$ the original ACS pixel scale).

The WFC3/IR data are from the {\it HST} Cycle 17 program GO-11563 (PI: G.
Illingworth), which aims at complementing the HUDF and the two HUDF05 parallel
fields \citep{oesch07} with WFC3/IR images in Y (F105W), J (F125W), and H
(F160W) of matching sensitivity, $\sim$29 mag \citep{bouwens10,oesch10}.  Here
we use only the first epoch of the images, released in September 2009, which
includes 18 orbits in Y, 16 orbits in J, and 28 orbits in H. We have carried
out our independent reduction of the raw data, and after rejecting images
affected by persistence in the J band, our final stacks reach 1$\sigma$ surface
brightness fluctuations of 27.2, 26.6 and 26.3 AB/\arcsec\ $^2$ in the three
bands, respectively, over an area roughly equal to the footprint of the WFC3/IR
camera (2.1 \arcsec\ $\times$ 2.1 \arcsec). We have drizzled the WFC3 images
from their original pixel size of 0.121\arcsec $\times$ 0.135\arcsec\ to
0.03\arcsec\ per pixel to match the scale of the GOODS and HUDF ACS images.
Further details on the production of the WFC3 data are given in
\citet{candelshst}. 

In addition to the {\it HST }/ACS and WFC3/IR images in the HUDF, the data used
in this paper also include panchromatic multi--wavelength photometry obtained
as part of the GOODS program, as the HUDF field is embedded in the GOODS south
field. The long wavelength baseline of the GOODS photometry enables us to
reliably select SFGs based on photometrically--derived stellar mass and
specific star formation rate (SSFR), while the deep {\it HST} optical and NIR
images allow us to obtain color maps of the galaxies with a resolution of $\sim
1$ kpc.

The GOODS south field has been observed with various telescopes and instrument
combinations, from the X--ray to the sub--millimeter and radio. Relevant to our
analysis here is the VLT/VIMOS ultra--deep U--band imaging \citep[]{nonino09},
as well as {\it HST}/ACS BViz \citep{giavalisco04}), VLT/ISAAC JHK,
Spitzer/IRAC 3.6, 4.5, 5.7, 8.0 $\mu$m, and Spitzer/MIPS 24 $\mu$m imaging.
Because the resolution of the images significantly changes from optical- to
IR-bands, we use an object template-fitting software dubbed TFIT
\citep{laidler07} to obtain matched multi--band photometry. The detailed
description of TFIT measured catalog of GOODS-S can be referred from \citet{dahlen10,ycguo12vjl}.

\subsection{Sample}
\label{data:sample}

In this work, we select SFGs with 1.5$<$z$<$2.5 in the HUDF to study their
clumps. Since we are measuring spatially resolved properties of galaxies
through photometry of only seven bands (ACS BViz and WFC3 YJH), in order to
reduce the number of free SED-fitting parameters, we restrict our sample to
only contain galaxies that have spectroscopic redshifts. The use of
spectroscopically observed galaxies reduces a major uncertainty of SED-fitting,
namely the uncertainty of (photometric) redshift. However, it also reduces the
size of our sample and might bias our sample towards the bright end. In the
HUDF, only 15 SFGs (SSFR $>$ ${\rm 0.01 Gyr^{-1}}$) have spectroscopic redshift
(spec-z) at 1.5$<$z$<$2.5 to enter our sample. We visually inspect the z-band
images of these galaxies and exclude five galaxies that have no obvious
multi-clump morphology. 

\begin{figure*}[htbp]
\center{\includegraphics[scale=0.9, angle=0]{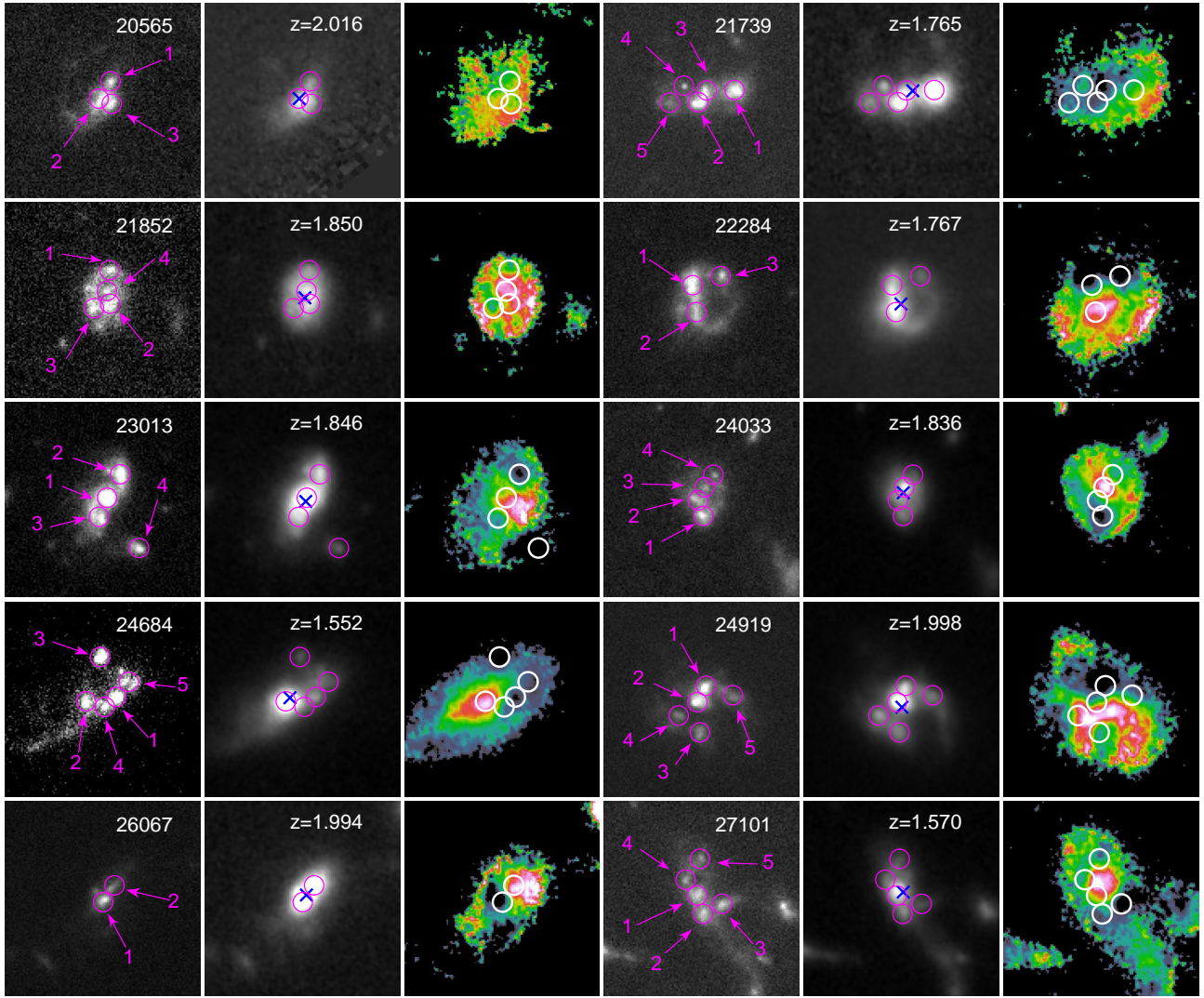}}
\caption[]{Montage of our 10 clumpy star-forming galaxies at z$\sim$2. Each row
shows images of two galaxies. For each galaxy, the panels from left to right
show the z-band, H-band and z-H maps. Their galaxy IDs are shown in their z-band
images, while redshifts in their H-band images. Small circles (magenta in the
z-band and H-band images, and white in the z-H maps) show the identified clumps. 
Blue ``X'' in the H-band images show the light-weighted centers. \\
\label{fig:mosaic}}
\vspace{-0.2cm}
\end{figure*}

About 67\% (10 out of 15 galaxies) of our sample are multi-clump systems in
their optical ACS z-band (rest-frame UV) images. This result is quite similar
to what \citet{elmegreen07} found, namely the majority of SFGs at z$\sim$2 is
clumpy galaxies. It is also consistent with the clumpy fraction that
\citet{wuyts12} measured through rest-frame UV light. However, we also note
that due to the use of spec-z, our sample is small and possibly biased toward
UV bright galaxies.  The bright UV emission may further suggest that our sample
is biased toward bluer \citep[e.g.,][]{vandokkum06,fs09} and more active
galaxies among all SFGs at z$\sim$2. Moreover, as shown by the recent work of
\citet{wuyts11b}, actively star-forming galaxies tend to have the largest sizes
\citep[see also, e.g.,][]{franx08}, which presumably introduces a further bias
in our sample toward large galaxies. To draw a robust conclusion on the
fraction of clumpy galaxies at z$\sim$2, a large sample (perhaps with
photometric redshift) covering a wide range of both luminosity and stellar mass
is needed. The ongoing 902-orbit {\it HST} Multi-cycle Treasure Program, Cosmic
Assembly Near-infrared Deep Extragalactic Legacy Survey
\citep[CANDELS][]{candelsoverview,candelshst} has already begun to
provide deep images over a large sky area to answer this question as well as to
provide robust statistics on clump properties, as demonstrated by
\citet{wuyts12}. We also anticipate that the present paper may lay the
groundwork for future studies in the CANDELS survey. The z-band, H-band, and
z-H mosaics of the 10 galaxies in our sample are shown in Figure
\ref{fig:mosaic}. The properties of the galaxies are shown in Table
\ref{tb:glx}.

Three galaxies in our sample are discussed in \citet{elmegreen05}: Galaxy
22284 (with ID 3484 in Elmegreen's paper), 21739 (Elmegreen ID 3465+), and
27101 (Elmegreen ID 6462+). These galaxies, classified as "clump-clusters" in
the paper, were chosen to study by Elmegreen et al. because of their large
angular size, good resolution into clump and interclump regions.
\citet{elmegreen05} used visual inspection on ACS i-band to identify clumps.
They identified more clumps in these galaxies than we find (see Table
\ref{tb:comp}). Although the different selection bands may contribute, the
different numbers of identified clumps in the two works could be mainly due to
different clump identification strategies.  We conservatively ask clumps to be
3$\sigma$ brighter than the diffuse components of the host galaxies (see the next
section), while \citet{elmegreen05} simply ask clumps to be bright compared to
the sky background. We will discuss the derived properties of clumps in the two
works in Sec. \ref{disc:other}.

%

\section{Clump Detection and Measurement}
\label{detection}

Clumps are detected from the HST/ACS z-band images through a hybrid of
automated detection and visual inspection.  In the image of each galaxy, a
clump is identified if a region contains at least six contiguous pixels that
are brighter than 3$\sigma$ confidence level of the average surface brightness
of the galaxy. If two or more sub-structures above 3$\sigma$ of the average
surface brightness of the galaxy are adjoining, we visually separate them and
determine the peak of each clump. The locations of each identified clumps are
shown by circles in the z-band images of galaxies in Figure \ref{fig:mosaic}.

Our choice of the ACS z-band as the detection band of clumps raises a
question: how do the number and nature of identified clumps depend on the
choice of the detection band? The question lies in two folds: (1) the impact of
the spatial resolution of images on clump identification and (2) the impact of
the use of different bands that are dominated by light of different stellar
populations on clump identification. This problem was highlighted by
\citet{fs11b}, who found that, for one galaxy in their sample, clumps
identified through WFC3 H-band and ACS i-band images have different physical
proprieties. To test the robustness of our clump identification through the
rest-frame UV band, we carry on the following two tests. First, we smooth the
ACS z-band images of clumps to match the resolution of WFC H-band and repeat
our identification procedure. Only one clump (\#4 of 24033) is missed, and
three pairs of clumps (\#2 and \#3 of 20565; \#2 and \#3 of 21739; and \#1 and
\#2 of 26067) are blended in the smoothed images. Given the fact that only 10\%
(4 out of 40) of our identified clumps are affected, we conclude that the
change of resolutions of images (from ACS to WFC3) would not have significant
impact on our clump identification and hence our later results. 

The second test is to identify clumps directly from WFC H-band images, which
sample the evolved stellar populations at z$\sim$2.  Besides suffering from the
impact of resolution, the H-band detection misses a non-negligible fraction,
$\sim$20\%, of clumps that are detected in the z-band. The missed clumps are
\#5 of 21739; \#4 of 23013; \#2 and \#4 of 24033; \#3 and \#4 of 24684; \#3,
\#4 and \#5 of 27101. These clumps are missed mainly because of lower contrast
of clumps against the background diffuse stellar population in the H-band, and
most of them are faint even in the z-band. Two clumps that are bright in the
z-band but not detected in the H-band are \#4 of 23013 and \#3 of 24684. An
important result in this test is that we do not find any new detected clumps
through using the H-band images. This is different from the case of
\citet{fs11b}, and can be attributed to the ultra-deep rest-frame UV images
used in our study, which allows us to probe even rest-frame UV faint (and hence
very red) clumps. Thus, we conclude that our clump identification through ACS
z-band images is superior to that through WFC H-band images, as the H-band
detected clumps are just a sub-sample of the z-band detected ones. Using the
z-band images, we are able to study the nature of clumps in a wide range of
physical properties.

For each identified clump, we measure its multi-band photometry from ACS BViz
and WFC3 YJH images, all smoothed to match the resolution of the WFC3 H-band
image ($\sim$0.15\arcsec). The details and tests done to ensure an accurate PSF
matching can be found in \citet{ycguo11peg}. The fluxes of clumps in each band
are measured within an aperture with size (diameter) of 0.3\arcsec, about two
times the FWHM of the H-band images. In order to separate the light of clumps
from that of diffuse components of galaxies, we subtract a diffuse background
from the flux of each clump. The surface brightness of the background is
calculated as the average flux per pixel of the host galaxy, after all clumps
in it (circles with diameter of 0.3\arcsec) are masked out. The product of the
surface brightness and area of a clump is then subtracted from the flux of the
clump. The diffuse background is subtracted for all clumps in all seven bands.

It is important to note that for all clumps in one galaxy, we subtract a {\it
constant} diffuse background across the whole galaxy. This subtraction might be
over-simplified, as the underling diffuse or ``disk'' component of a galaxy is
likely to have a non-constant profile (e.g., an exponential disk profile). In
that case, we would over-subtract background for clumps in outskirts of the
galaxy, while under-subtract it for those close to the center. We argue that
this over-simplified subtraction would not significantly change our results on
the comparison of clump properties and ``disk'' properties, as on average, only
15\% of the raw flux of each clump is subtracted. However, it may induce false
signals into our studies on the radial variation of clump properties across the
host galaxies. We will discuss its influence later (Sec. \ref{prpvar:bkgsub}).
We also note that the amount of subtraction is comparable to the photometric
uncertainty that is caused by our use of the arbitrary $d$=0.3\arcsec\
aperture. When we enlarge or shrink our aperture size by 2 pixels, the typical
change on flux is about 15\%$\sim$20\%. Therefore, we use the difference of the
fluxes measured at $d\pm0.06$\arcsec\ as the uncertainty of fluxes of clumps.

The physical properties of clumps are measured through fitting seven-band SEDs
to stellar population synthetic models. We use the updated version (CB09,
Charlot \& Bruzual in prep.) of \citet{bc03}, which includes a new treatment on
the contribution of TP-AGB stars, as our stellar population library. The
initial mass function (IMF) of Salpeter \citep{salpeter55} is used. The
redshifts of all clumps are fixed at the spec-zs of their host galaxies during
the fitting.  Since the SFH of galaxies at z$\sim$2 is controversial
\citep[e.g.,][]{joshualee09,joshualee10,maraston10,papovich11}, we fit each
clump with three different types of SFHs: exponential declining ($\tau$-model),
constant (CSF), and exponential increasing. Among the three fits of each clump,
the one with the smallest reduced $\chi^2$ is chosen as the best-fit model and
its corresponding parameters are used as the physical properties of the clump.
For each fitting parameter, we use the standard deviation of the three best-fit
values from different SFHs as the uncertainty of the parameter, as we believe
that the largest uncertainty of SED-fitting is rooted from the uncertainty of
SFHs. We fix the metallicity to the solar value for all clumps. We will discuss
the influence of metallicity variation among clumps on our results later (Sec.
\ref{prpvar:metallicity}). When deriving the physical properties from the
best fit template, we adopt the following definitions. To measure SFR, we
average the SFH of the best fit template over the last 100 Myr, which allows a
meaningful comparison between SFR derived through SED-fitting and that through
rest-frame UV continuum empirically. The age of objects in our paper
corresponds to the time from the onset of their star formation to their
redshifts. And our stellar mass measurement accounts for the loss of stellar
mass over time.

We also measure the fluxes and properties of the diffuse component (called
``disk'' thereafter) in each galaxy. The flux of a ``disk'' is measured by
subtracting the total fluxes of clumps in its host galaxy from the flux of the
galaxy. The physical properties of ``disks'' are measured with the same method
that is used for clumps. We note that whether these diffuse components are real
disks or not is still unknown. Actually, it is a key to understand the
formation mechanisms of clumps. The existence of underlying disks is a
necessary condition for the scenario of fragmentation due to gravitational
instability. Kinematic studies, e.g., \citet{fs09}, are needed to understand
the nature of the diffuse components. In this paper, we just call them
``disks'' for simplicity.

The properties of clumps and ``disks'' are shown in Table \ref{tb:clump}.

\section{Physical Properties of Clumps}
\label{prp}

\subsection{Rest-frame UV--Optical Color of Clumps}
\label{prp:color}

To understand the nature of our clumps, we study their positions in
color--magnitude diagram (CMD). In CMD, present-day galaxies are well separated
into two populations: red-sequence and blue cloud
\citep[e.g.,][]{blanton03b,bell04}.  This color bimodality is observed to exist
up to z$\sim$1.5 \citep[e.g.,][]{bell04,brammer09,mendez11}.  The red-sequence
is believed to consist of old and quiescent galaxies, while the blue cloud
consists of young and active galaxies.  The locations of clumps in CMD, namely
among red-sequence or blue cloud, would shed a light on whether they are as
quiescent as local red spheroids/bulges or still actively forming stars.

\begin{figure}[htbp]
\center{\includegraphics[scale=0.8, angle=0]{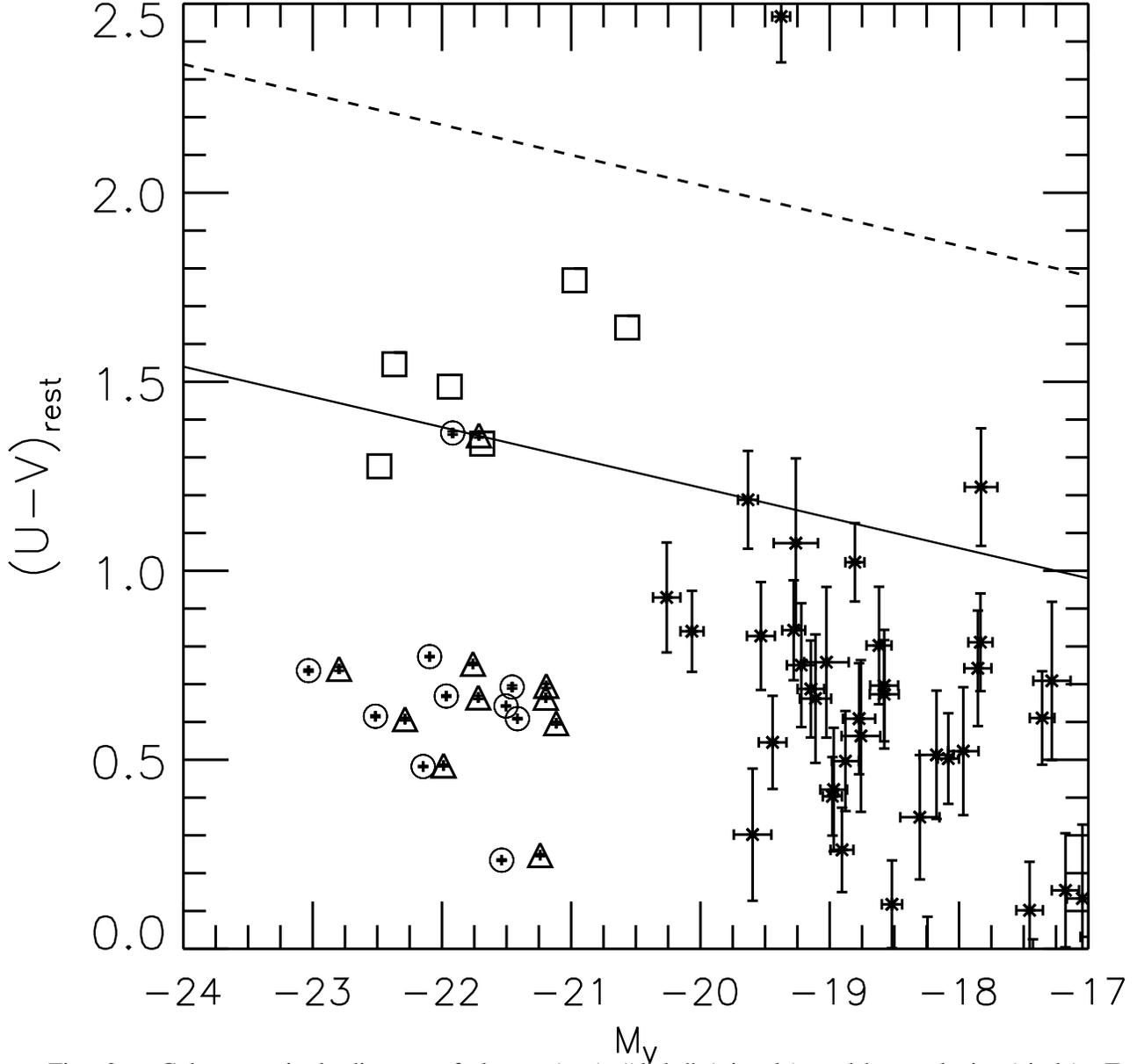}}
\caption[]{Color--magnitude diagram of clumps (star), ``disks'' (triangle), 
and host galaxies (circle). Together plotted are six passively-evolving 
galaxies at z$\sim$2 (squares) from \citet{ycguo11peg}. We also plot
the separation of red-sequence and blue cloud at z=0 from \citet{bell04}
(dashed line). The solid line is the extrapolation of the separation 
from z=0 to z=2 by using the formula of evolution of the separation of 
\citet{bell04}. \\
\label{fig:bimodal}}
\vspace{-0.2cm}
\end{figure}

Figure \ref{fig:bimodal} shows the diagram of rest-frame (U-V) color vs.
absolute rest-frame V-band magnitude for clumps (starred symbols). Rest-frame
magnitudes are obtained through interpolating the observed seven-band
photometry to corresponding rest-frame wavelength, based on the redshift of
each clump. In the figure, we also plot the ``disks'' (triangles) and host
galaxies as a whole (circles). We also plot the six passively-evolving galaxies
from \citet{ycguo11peg} (squares) in the figure. The separation of red-sequence
and blue cloud shifts blueward as the redshift increases, because the cosmic
age decreases with redshift. To obtain a red-blue separation line for z$\sim$2,
we extrapolate the separation line at z=0 of \citet{bell04} to z=2 with the
empirical evolving formula proposed in their paper:
$<U-V>=1.23-0.4z-0.08(M_V-5logh+20)$. The red--blue separation lines at z=0 and
z=2 are shown by dashed and solid lines in the figure.

In general, clumps are as blue as their surrounding ``disks'' and host
galaxies. Only few clumps reach above the red--blue sequence at z=2 and might
have properties similar to quiescent galaxies at z$\sim$2. Overall, the blue
color indicates that the star formation activity is still strong and has not
been widely quenched in clumps. We note that our clumps are identified
through the excess of the rest-frame UV light in the diffuse background.
Therefore, we expect them to have {\it bluer} color than their surrounding
``disks''. However, Figure \ref{fig:bimodal} does not support such expectation.
What's more, it shows that, compared with ``disks'' or host galaxies, clumps
seem to have broader color dispersion. This broader scatter suggests that our
clumps cover a wide range of physical properties (e.g., age or extinction),
have diverse star formation histories, or are in different evolution stages.
This result seems contradictory to the simple ``Christmas tree'' model, in
which clumps are rapidly disrupted once they have formed. In that case, the
properties of clumps may concentrate within a narrow range. We will quantify
the dispersion of clump colors in Sec. \ref{prp:cvd}.

\subsection{SFR--Stellar Mass Relation of Clumps}
\label{prp:sfr}

\begin{figure}[htbp]
\center{\includegraphics[scale=0.8, angle=0]{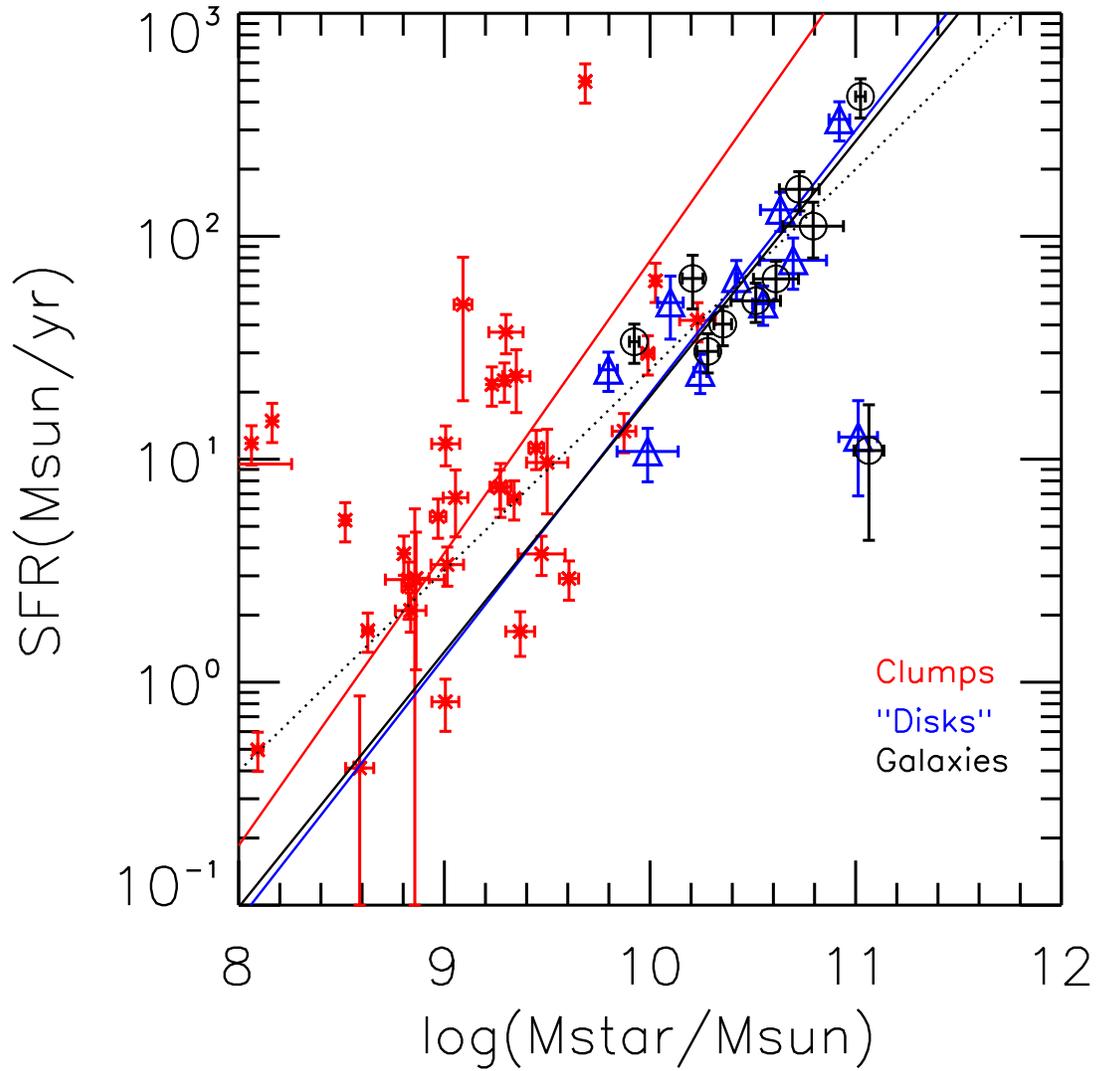}}
\caption[]{SFR--stellar mass relation for clumps (star), ``disks'' (triangle),
and host galaxies (circle). The SFR--stellar mass relation of each population
is fitted by a straight line (red, blue, and solid black for clumps, 
``disks'', and host galaxies, respectively). The dotted line is the relation 
of \citet{daddi07a} for BzK galaxies. \\
\label{fig:sfr}}
\vspace{-0.2cm}
\end{figure}

Figure \ref{fig:sfr} shows the SFR--stellar mass relation of clumps (red stars)
as well as that of ``disks'' (blue triangles) and whole galaxies (black
circles).  SFR and stellar mass are measured through SED-fitting with various
assumed SFHs as described in Sec. \ref{detection}. The best linear fit for the
relation of each population is also shown in the plot. In order to
evaluate the significance of the SFR--stellar mass relation of both clumps and
``disks'', we run bootstrapping 100 times to randomize the pair of SFR and
stellar mass of clumps and disks and calculate the correlation coefficients
(defined as ${\rm \sqrt{COVAR[X,Y]/(VAR[X]*VAR[Y])}}$, where COVAR is
covariance and VAR is variance of X and Y, and X and Y stand for stellar mass
and log(SFR)) for both the observed and randomized relations.
By comparing the correlation coefficients of both the observed and randomized
relations, we find the relation of both clumps and ``disks'' are significant at
at least 3$\sigma$ level. The SFR--stellar mass relation of clumps and
``disks'' have almost the same slope, although the two populations occupy
different ends of the stellar mass range: the stellar masses of clumps spread
over the range between ${\rm 10^8 M_\odot}$ and ${\rm 10^{10} M_\odot}$, while
those of ``disks'' between ${\rm 10^{10} M_\odot}$ and ${\rm 10^{11} M_\odot}$.
However, the normalization of the relation of clumps is about five times higher
than that of ``disks'', converting to a higher SSFR of clumps. The higher SSFR
is consistent with the spectroscopic measures of \citet{genzel11}. 

The reason of the higher SSFR of clumps is important for understanding the
nature of clumps, but unfortunately still unknown. It could be due to the fact
that the locations of clumps sample the very high end of the gas density
distribution of host galaxies. As a sequence, the gas--to--stellar mass ratio
is higher in clumps than in ``disks''. This case is likely true, as these
clumps are believed to form through gravitational instability in the gas-rich
"disks". Such process indicates a high gas density in their birth locations.
However, based on our later analysis, the (projected) stellar mass density of
clumps is also higher than that of "disks". Therefore, it is still uncertain
whether the gas--to--stellar mass ratio of clumps is comparable for that of
``disks''. If they are comparable, clumps would have higher efficiency to
convert gas into stars to yield higher SSFR. In this case, other mechanisms,
for example a different star-formation law, should be used to explain the
higher star formation efficiency of clumps. Further observations on spatially
resolved gas density (e.g., through ALMA) are required to investigate the star
formation activity in clumps.

Figure \ref{fig:sfr} also shows that the SFR--stellar mass relation (or SSFR)
of the galaxies as a whole has almost the same slope and normalization as that of
``disks''.  It suggests that ``disks'' contribute the majority fraction of SFR
and stellar mass of the host galaxies.  Clumps, on the other side, stand out as
regions with enhanced SSFR in the ``disks''.

\subsection{Contribution of Clumps to Host Galaxies}
\label{prp:cvg}

\begin{figure*}[htbp]
\center{\includegraphics[scale=0.4, angle=0]{./fig4.ps}}
\caption[]{Fractional contributions of clumps to their host galaxies in terms
of rest-frame U-band luminosity (top left), V-band luminosity (top right),
stellar mass (bottom left), and SFR (bottom right). Solid histograms show the
distributions of the contribution of each clump, while dotted lines show the
distributions of the total contribution of all clumps in each host galaxy.
The horizontal error bar in each panel shows the typical error of each
measurement. \\
\label{fig:cvg}}
\vspace{-0.2cm}
\end{figure*}

Figure \ref{fig:cvg} shows the fractional contributions of clumps to their host
galaxies, in terms of rest-frame U-band and V-band luminosity, stellar mass and
SFR. We show the distribution of the contributions of both each individual
clump (solid line) and the sum of all clumps in one galaxy (dashed line). For
the U-band and V-band luminosity (top left and top right panels), the
individual contribution of clumps runs from 1\% to 10\%, with a median of
$\sim$5\%, while the total contribution of all clumps sharply peaks around
20\%.  There is no obvious difference between the clump contributions to both
luminosities. The clump contribution to stellar mass (bottom left) is similar
to that of the light, with a broad distribution of individual contribution
running from 1\% to more than 10\%, and a concentrated total contribution
peaking around 20\%. We note that the similar contributions to stellar
mass and to light of individual clumps do not imply a similar M/L ratio among
clumps. Actually, the ${\rm M/L_{V}}$ ratio of clumps covers a wide range, from
$\sim$0.05 ${\rm M_\odot/L_{\odot,V}}$ to several ${\rm M_\odot/L_{\odot,V}}$,
with the median of 0.5 ${\rm M_\odot/L_{\odot,V}}$ and the standard deviation
of 0.5 dex. The widespread M/L ratio of clumps is consistent with our previous
argument that clumps are found at different evolutionary stages.

The contribution of clumps to SFR (bottom right) is higher than that to light
and stellar mass. The individual contribution peaks around 10\%, while the
total contribution peaks around 50\%. The low contribution to stellar mass and
high contribution to SFR are consistent with the fact the clumps have enhanced
SSFR relative to ``disks'' or galaxies as a whole.

\subsection{Clumps vs. ``Disks''}
\label{prp:cvd}

\begin{figure}[htbp]
\center{\includegraphics[scale=0.4, angle=0]{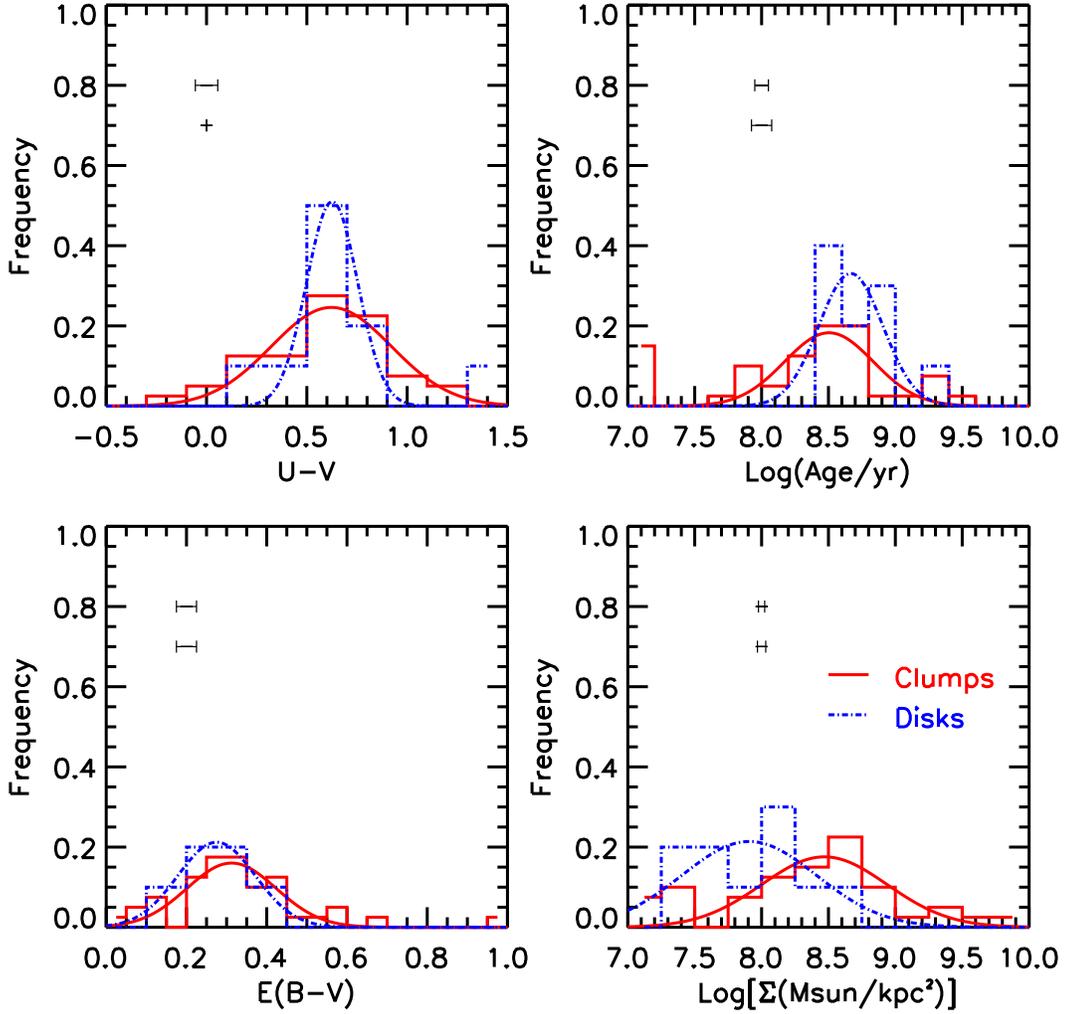}}
\caption[]{Comparisons between physical properties of clumps and their
surrounding ``disks''. Red histograms show the distributions of properties of
clumps, while blue those of ``disks''. Red and blue curves show the Gaussian fits
to the red and blue histograms. Typical measurement errors are shown as
horizontal error bars (upper for clumps and lower for ``disks'') in each panel.
\\
\label{fig:cvd}}
\vspace{-0.2cm}
\end{figure}

Figure \ref{fig:cvd} shows the difference between clumps and ``disks'' in terms
of the distributions of UV--optical colors (top left), ages (top right), dust
extinction (bottom left) and projected stellar mass densities (bottom right) of
clumps (red) and ``disks'' (blue). Each distribution is fitted by a Gaussian
function.  This figure shows that the mean UV--optical color of clumps is 
similar to that of disks (U-V$\sim$0.6). However, the color distribution of
clumps is broader than that of disks. The standard deviation of the former is
0.30, while that of the latter is 0.12. Clumps can be as red as U-V$\sim$1.2
and as blue as U-V$\sim$-0.2, while ``disks'' are concentrated within
0.2$<$U-V$<$0.8, except for one ``disk''.  The broader color distribution of
clumps can be attributed to either different SFHs among clumps or different
evolution stages of clumps. 

Our SED-fitting method only provides limited information on SFHs, however, the
distribution of ages of clumps gives us a hint that the broader color
distribution of clumps is more likely associated with the evolution stages.  On
average, the ages of clumps are only slightly (0.2 dex) younger than that of
``disks''. However, the ages of clumps have a broader distribution, covering
the range from 0.01 Gyr to a few Gyr, with the standard deviation of 2.1 dex.
In contrast, the ages of ``disks'' are concentrated within 0.3 to 1 Gyr, with a
standard deviation of 1.7 dex. Moreover, in Sec. \ref{prpvar}, we will show
that both color and age of clumps change with the galactocentric distance of
clumps.  Based on the trend of their radial variations, one can expect a loose
correlation between colors and ages of clumps, which would support our
speculation that the broader color distribution of clumps is due to the broad
age distribution (and hence different evolution stages). However, we note
that the age determination in SED-fitting is not robust and strongly depends on
the assumed SFHs. As shown by the horizontal error bars in the top right panel,
the uncertainty of age in our study, namely the standard deviation of the ages
measured by SED-fitting with three different SFHs, is about 0.5 dex and 0.3 dex
for clumps and ``disks'' respectively, larger than the difference between the
mean age of the two populations.

We also argue that other factors, such as extinction and metallicity, are not
likely to be the major contributor of the broader color distribution of clumps.
The difference of the E(B-V) distributions of clumps and ``disks'' is not
significant, as can be seen from the bottom left panel of Figure \ref{fig:cvd}.
The mean and standard deviation of the E(B-V) distribution of clumps are 0.31
and 0.11, while those of ``disks'' are 0.27 and 0.10. The differences of both mean
and standard deviation of the two components are actually smaller than the
typical uncertainty of our E(B-V) measurement ($\sim$0.05). Although a few
clumps do have very high dust extinctions, the insignificant difference between
the E(B-V) distributions of clumps and ``disks'' suggests that extinction is
not a major contributor of the broader color distribution of clumps. Another
possible reason of the broad color distribution of clumps is the metallicity
variation among clumps. Since metallicity is not a free parameter in our
SED-fitting, we cannot draw conclusions on the metallicity distribution of
clumps. Instead, we will discuss the effect of metallicity variation in Sec.
\ref{prpvar:metallicity}.

A more significant difference between clumps and ``disks'' comes from the
distributions of their projected stellar mass densities. The high resolution of
ACS z-band images allows us to resolve each clump and hence measure its projected
stellar density. By comparing the FWHM of clumps (after subtracting diffuse
background) to that of ACS z-band PSF, we find that all our clumps except one
are (marginally) resolved in the ACS z-band image. The fraction of resolved
clumps in our sample is higher than that in \citet{fs11b}, who found 18 out of
27 of their NIC2 clumps are resolved. The higher fraction of resolved clumps in
our study can be explained by the higher resolution of ACS z-band image
(0.12\arcsec), compared to that of NICMOS NIC2 images (0.15\arcsec). However,
we also caution that the measurements of the projected stellar density of
clumps in our study may suffer from systematical uncertainty, because the size
of clumps is not precisely determined. To do so requires a detailed
investigation on the light profiles of clumps and their host galaxies
simultaneously. In this work, instead, we simply use the size of the apertures
that we use to measure photometry as the size of clumps. The size is about two 
times the FWHM of H-band PSF, equivalent to about 5$\sigma$ of the PSF, and is
believed to be a good representative of clump sizes. The projected stellar
density measured with this size can be treated as an average density of a
clump.

The bottom right panel of Figure \ref{fig:cvd} shows that clumps are on average
eight times denser than ``disks''.  This result is non-trivial, because we identify
clumps based on z-band images, which, observing the rest-frame UV light, is
more sensitive to star formation than to stellar mass. Identified as prominent
sub-structures in the z-band images, these clumps are expected to have active
star formation but not necessarily to be denser. Using the light from both
rest-frame UV and optical bands, our SED-fitting method turns out to show that
clumps are regions with not only enhanced SSFRs but also more concentrated
stellar distributions.  

The higher stellar surface densities of clumps are consistent with the
hypothetical scenario that clumps are formed through gravitational
instabilities
\citep[e.g.,][]{noguchi99,immeli04a,immeli04b,bournaud07,bournaud08,elmegreen08,dekel09,ceverino10}.
\citet{genzel11} calculated the maps of Toomre {\it Q}-parameter for four
galaxies in their sample and found that throughout the outer disks and toward
the clumps, the {\it Q}-parameter is at or even significantly below unity.  The
small ($<$1) {\it Q}-parameter is evidence of gravitational instability. Since
the total {\it Q}-parameter is inversely proportional to the sum of the
molecular gas and stellar surface densities (see 
Eq. (2) of \citet{genzel11}), the regions with low {\it Q}-parameter (namely
regions toward clumps) should have higher surface densities of gases and/or
stars, as our results show.
However, we remind that the possibility that the instability is driven by other
violent processes, such as interaction or merger, cannot be ruled out simply
based on the estimation of {\it Q}-parameter. 

\section{Clumps across Host Galaxies}
\label{prpvar}

In a widely held view, clumps are expected to migrate toward the gravitational
centers of their host galaxies due to dynamical friction against the
surrounding disks or clump interactions and eventually coalesce into a young
bulge in several dynamical timescales ($\sim$0.5 Gyr) to form the progenitor of
today's bulges. Alternatively, they could also be disrupted by either tidal
force or stellar feedback to form part of a thick disk
\citep[e.g.,][]{escala08,dekel09}.  If clumps are able to survive the several
dynamical timescales of migration, they are expected to exhibit a broad age
dispersion, with older clumps generally closer to the galactic centers.
Moreover, other physical properties of clumps would also change as the clumps
migrate toward the galactic centers. For example, the gas outflows, as observed
by \citet{genzel11}, would be sufficiently strong to expel a large fraction of
gas of clumps so that clumps are expected to become less efficient at forming
stars when sinking toward centers. In this section, to understand the evolution
of clumps, we study the radial variations of colors and physical properties of
clumps along their host galaxies. We also discuss the effect of diffuse
background subtraction (see Sec. \ref{detection}) on our results. We will also
discuss the possibility of metallicity variation as an explanation of the
observed color radial variation. 
 
\subsection{Radial Variation of Color}
\label{prpvar:colorvar}

\begin{figure}[htbp]
\center{\includegraphics[scale=0.5, angle=0]{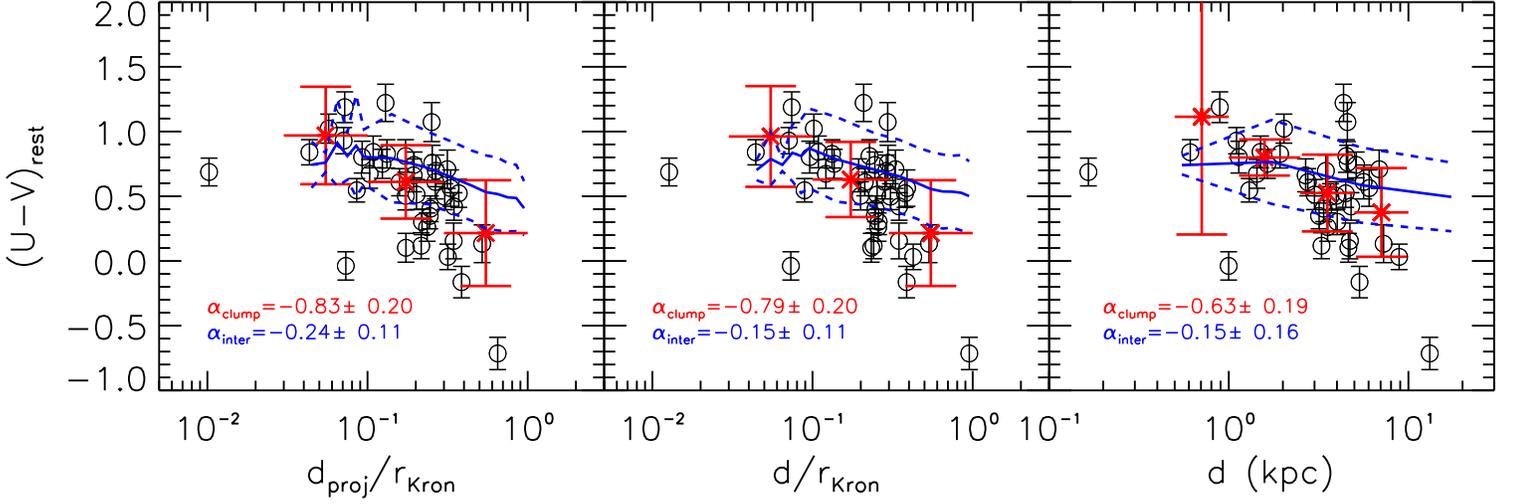}}
\caption[]{Radial variation of the rest-frame U-V color of clumps as a function
of their galactocentric distances to the H-band light-weighted centers of
their host galaxies. The distance is calculated in different ways in the three
panels. {\it Left}: The distance is the projected distance, scaled by the
H-band Kron radius of the host galaxy. {\it Middle}: It is the deprojected
distance, scaled by the H-band Kron radius of the host galaxy. {\it Right}: It
is the deprojected physical distance, in unit of kpc.  To calculate the
deprojected distance, we assume a circle configuration with the inclination
angle equal to the axis ratio for each galaxy. In all panels, each circle with
error bar stands for one clump. The red stars and red vertical error bars show
the mean and 1$\sigma$ deviation (after 3$\sigma$-clipping) of the clumps in
each distance bin, while the bin size is shown by the red horizontal error bar.
The solid and dashed blue lines in each panel show the mean and standard
deviation of color variation of interclump pixels of all galaxies. The best-fit
slope of the color gradients of clumps and interclump pixels are given in 
each panel.
\\
\label{fig:colorvar}}
\vspace{-0.2cm}
\end{figure}

Figure \ref{fig:colorvar} shows the radial variation of the rest-frame U-V
color of clumps across the host galaxies. We use the H-band 
light-weighted centers of host galaxies to represent the galactic centers, as
the H-band is closest to the peak of stellar emission in all available bands.
We calculate the galactocentric distance of each clump in the following
three ways and study the radial variation in each case, respectively: (1)
projected distance, scaled by the H-band Kron radius of its host galaxy (left
panel); (2) deprojected distance, scaled by the Kron radius (middle panel); and
(3) deprojected physical distance, in unit of kpc (right panel). All panels
show a clear trend that clumps close to the centers of their host galaxies
have redder rest-frame UV/optical colors than those in the outskirts. 
Although the slope of the radial color variation (or color gradient), defined
as $\alpha = \Delta (U-V) / \Delta Log(R)$, varies with the definition of the
galactocentric distance, it is significant beyond the 3$\sigma$ confidence
level in all cases (the values of the slope can be read from Figure
\ref{fig:colorvar}). Therefore, we conclude that the radial color variation is
an intrinsic feature of clumps and not affected by the distance definition. In
later analysis, we use the scaled projected distance, as it gives us the
strongest signal of radial variation and is independent of the assumption of
the circle configuration of galaxies, which is used in our calculation of the
deprojected distance, but still very uncertain for our galaxies.  With the
projected distance, the average colors of clumps become redder by 0.8 mag from
radius of 0.7 ${\rm r_{Kron}}$ to 0.07 ${\rm r_{Kron}}$. This picture is
broadly consistent with the scenario of inward migration of clumps.  However,
since the color is governed by a few factors: age, dust extinction and
metallicity, the radial variation of color itself only provides an indirect
comparison with theoretical hypothesis. The radial variations of physical
properties are needed to make a direct comparison with models.

There is one more issue that may also complicate the inference of the
nature and fate of clumps from color variation and invalidate the inward
migration scenario. It is the radial color variation of the diffuse components
(``disk'') of our galaxies.  If the ``disks'' exhibit the same radial color
variation (or gradient), 
it is then very likely that clumps are formed and, after that, stay in the
locations where they are observed today. In this case, the radial variation of
clumps can be explained by the radial variation of their host ``disks'',
because thus formed clumps would have the same dust extinction and/or metallicity
as that of their birthplaces in the host ``disks''. In order to examine the
possible color gradient of host ``disks'', we measure the rest-frame U-V color
of all interclump pixels (defined as pixels that are 0.3\arcsec\  away from the
center of any clumps) and study their radial variation. The mean and standard
deviation of the color of interclump pixels are shown as a function of
galactocentric distances in Figure \ref{fig:colorvar}. Interclump pixels show
the similar trend of color gradient as clumps: red in center and blue in
outskirts. However, the slope of the color gradient of interclump pixels
($\alpha_{inter}$) is significant at only 2$\sigma$ level with the rescaled
projected galactocentric distance, and decreases to even only 1$\sigma$ level
with the deprojected physical distance, indicating a mild color gradient.
More importantly, $\alpha_{inter}$ is significantly, at least at 3$\sigma$
level, larger (flatter) than $\alpha_{clump}$, the slope of the color gradient
of clumps. The values of $\alpha_{inter}$ and $\alpha_{clump}$ can be read from
Figure \ref{fig:colorvar}. The significant difference between the two slopes
suggests that (1) the color gradient of ``disks'' can only explain a small part
of the color gradient of clumps and (2) the formation and evolution of clumps
are somehow dynamically separated from those of ``disks''. Therefore, a
mechanism such as the inward migration is needed to explain the steeper
gradient of clumps. 

\subsection{Radial Variations of Physical Properties} 
\label{prpvar:prpvar}

\begin{figure}[htbp]
\center{\includegraphics[scale=0.4, angle=0]{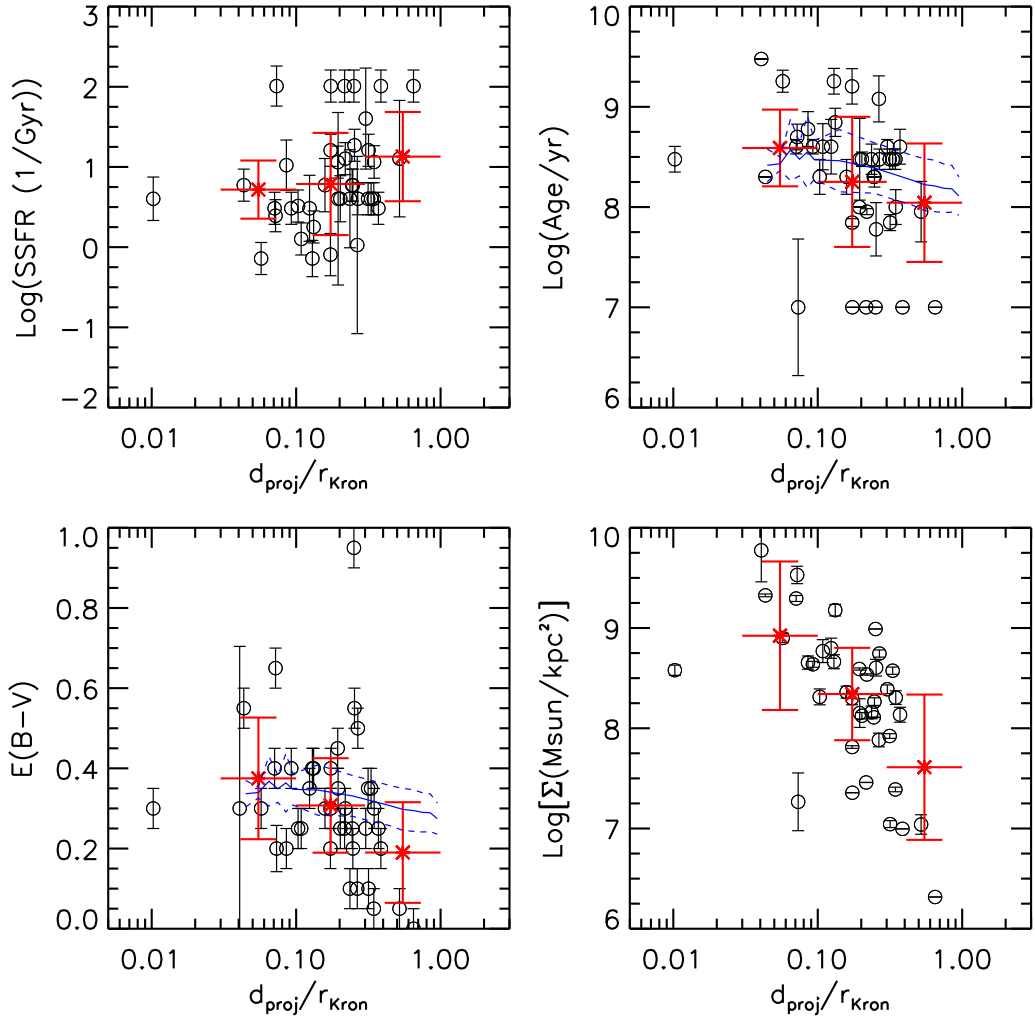}}
\caption[]{Radial variation of physical properties of clumps as a function of
galactocentric distance. Symbols and colors are same as those in Figure
\ref{fig:colorvar}. \\
\label{fig:prpvar}}
\vspace{-0.2cm}
\end{figure}

Figure \ref{fig:prpvar} shows the radial variations of SSFR (top left), age
(top right), E(B-V) (bottom left) and stellar surface density (bottom right) of
clumps. Same as in Figure \ref{fig:colorvar}, the galactocentric distances
of clumps are scaled by the H-band ${\rm r_{Kron}}$ of host galaxies. Clumps
close to the centers of host galaxies have lower SSFRs, older ages, higher dust
extinctions and higher stellar surface densities. On average, clumps located at
${\rm d=d_{proj}/r_{Kron}<0.1}$ have four or five times lower SSFR than those at
${\rm d>0.5}$. The average age of clumps at ${\rm d>0.5}$ is about 100 Myr,
while that of those at ${\rm d<0.1}$ is about 700 Myr. The trend of E(B-V) is
mild, increasing only 0.2 from ${\rm d>0.5}$ to ${\rm d<0.1}$.  The stellar
surface density changes dramatically, increasing about 25 times from ${\rm
d>0.5}$ to ${\rm d<0.1}$.

As we discuss in Sec. \ref{prpvar:colorvar}, the host ``disks'' show a
mild color gradient, which may contribute a small fraction of the observed
radial color variation of clumps. Here we try to examine the implication of
this color gradient on the possible gradient of physical properties of
``disks''. In order to do so, we have to run SED-fitting in each interclump
pixel. However, the large photometric uncertainties of interclump pixels,
especially of those in HST/ACS bands due to the relative low rest-frame UV
emission of the diffuse components (as one can infer from Figure
\ref{fig:mosaic}), prevent us from getting reliable fitting results for
individual interclump pixels. In our study, we take a detour by assuming the
(U-V) color--extinction relation and the color--age relation of each interclump
pixel follow those of the integrated ``disks''. Thus, we extrapolate these
relations of ``disks'' to the color of individual interclump pixels to obtain
an estimation of the extinction and age of the pixels. We believe that such
extrapolation is valid because the ``disk'' is the integration of all
interclump pixels and hence represents the average color--extinction and
color--age relations of the pixels.

The age and extinction gradients of interclump pixels are shown in Figure
\ref{fig:prpvar}. Similar to the case of color gradient, both gradients are
mild and cannot fully explain the observed radial variation of clump
properties. The age gradient of interclump pixels varies only from $\sim$300
Myr at the projected galactocentric distance of $\sim$0.07 ${\rm r_{Kron}}$ to
$\sim$100 at $\sim$1 ${\rm r_{Kron}}$, while a large fraction of clumps are
found to have age $>$700 Myr at $\sim$0.07 ${\rm r_{Kron}}$ or to have age
$<$100 Myr at $\sim$1 ${\rm r_{Kron}}$. The extinction of interclump pixels
varies only by ${\rm \Delta E(B-V)} \sim$ 0.1 throughout the ``disks'', while
that of clumps varies by ${\rm \Delta E(B-V)} \sim$ 0.2 from galactic centers
to outskirts. Although not surprising, these results again demonstrate that the
observed radial variations (or gradient) of clump properties are not a simple
reflection the gradient of ``disks''. Clumps have their intrinsic radial
variations so that their formation and evolution are separated from those of
``disks'', as discussed in Sec. \ref{prpvar:colorvar}. 

The trends of these physical properties are consistent with the scenario of
clumps migrating toward galactic centers in timescales of $\lesssim$1 Gyr.  The
trend of SSFR indicates that if the scenario is true, the intensity of star
formation activity in clumps reduces as they migrate toward the gravitational
centers. The reason of lower SSFR toward centers could be the gas outflow from
massive clumps \citep{genzel11}. However, the outflow is not strong enough to
fully quench the star formation in clumps, because even for clumps that are
closest to galactic centers, their SSFRs are still several ten times higher
than the usually quoted value for quiescent galaxies (${\rm 10^{-2} yr^{-1}}$).

The trends of age and E(B-V) are actually coupled due to the age--extinction
degeneracy. Both high extinction and old age can be used to explain the
relative red color of central clumps. To break the degeneracy requires resolved
rest-frame NIR images, which unfortunately are not available to date. In our
results, both parameters are partly responsible for the relative red color of
central clumps. This shared responsibility could be real or just a reflection
of the degeneracy. If the latter, using a sole parameter to explain the trend
of U-V colors of clumps requires the trend of the parameter being even more
prominent than what we see in the figure.  That is, central clumps could be
even older or more obscured. However, neither of the two situations is likely
to be true. First, the stellar feedback would expel gas out of clumps as well
as disrupt dust grains. The timescale of disrupting dust grains is much shorter
than that of expelling gas and quenching star formation \citep{draine09}. As a
result, it is unlikely to find clumps with very high dust extinctions
(E(B-V)$>$0.6) but low SSFRs toward the gravitational centers of host galaxies.
Second, the ages of central clumps are already around 1 Gyr in the figure. A
more prominent age trend would result in an average age of 2 Gyr or even older.
It is hard to explain why these clumps were formed so early but have not fully
migrated into the centers of galaxies to form bulges. Therefore, we conclude
that both age and extinction should be responsible for the relative red color
of central clumps.  

\subsection{Effect of Diffuse Background Subtraction}
\label{prpvar:bkgsub}

As described in Sec. \ref{detection}, we subtract a {\it constant} diffuse
background for all clumps in each galaxy. The real diffuse or ``disk''
component of a galaxy, however, is unlikely to have a constant profile, but
instead to have a non-constant (e.g., exponential disk) profile. The simplified
assumption of {\it constant} background would result in an over-subtraction for
clumps in outskirts and an under-subtraction for central clumps. This problem
exists for all bands but is more sever for red bands (e.g., H-band), since the
background--clump contrast in the red band images is lower than that in blue
band images, as one can infer from the z-band and H-band images in Figure
\ref{fig:mosaic}. Therefore, the imperfect subtraction on diffuse background
would redden the color of central clumps and blue the color of clumps in
outskirts, inducing a false signal on the trend of radial variation of color in
Figure \ref{fig:colorvar} and subsequently on the trend of radial variations of
physical properties in Figure \ref{fig:prpvar}.

Accurate subtraction of diffuse background is complicated. It requires
knowledge on the light profile of underlying ``disk'' components of host
galaxies. The commonly used method to obtain the light profile is fitting
galaxy image with a \sersic profile. Unfortunately, the image fitting technique
has a few shortcomings: model dependent, not suitable for asymmetric source,
and affected by the existence of clumps, which make it problematic for
determining the underlying components of high redshift irregular-like clumpy
galaxies in our sample. \citet{elmegreen09a} neglected the contribution of
diffuse background when they studied the clumps in HUDF. This seems reasonable
for their study, because they only used HST/ACS optical images, where the
background--clump contrast is high so that the flux of clumps is 2--4 times
higher than their surroundings. 

In another study, \citet{fs11b} subtracted background from their NIC2
H-band images of clumpy galaxies and explored the impact of different
background-subtraction schemes, including the one of no subtraction, on their
results. They tracked the light profile of clumps until an upturn or a break
appears and then used the surface brightness of the upturn or break as the
surface brightness of local diffuse background. For clumps without an upturn or
break, they measured the background just outside the photometric apertures of
clumps. This method is sensible, but somehow subject to the determination of
the upturn or break in a smoothly changing light profile, which may vary from
person to person. Moreover, subtracting values from the upturns or from pixels
just outside the clump photometric apertures would result in an
over-subtraction, because (1) the upturn is more likely to be caused by the
overlapping of two nearby clumps rather than by the domination of background
and (2) there is no reason that the background would immediately dominate the
flux just outside the photometric apertures of clumps. This possible
over-subtraction could partly explain their findings that the local background
light is typically 3--4 times higher than the background-subtracted clump
fluxes, while in our study, the background only accounts for on average a few
tens percent of the raw clump fluxes, with few of $\gtrsim$50\%. 
\citet{fs11b} also estimated the uncertainty on the clump light contributions
due to background subtraction by comparing the background-subtracted results
with two other measurement methods: directly PSF measurement and raw photometry
measurement without subtraction. They found that the clump light contributions
are uncertain to a factor of $\sim$3. More importantly, one of their intriguing
results, namely the trends of redder colors and of older ages for clumps at
smaller galactocentric radii, is not significantly changed by using either
background-subtracted or raw photometry. 

\begin{figure}[htbp]
\center{\includegraphics[scale=0.8, angle=0]{./fig8.ps}}
\caption[]{Radial variation of the rest-frame U-V color of clumps as a function
of their galactocentric distances, under different schemes of diffuse
background subtraction. Small filled blue triangles show the case of {\it
local} background subtraction, while red circles show that of {\it zero}
background subtraction. Large empty blue triangles (red circles) with error
bars show the mean and 1$\sigma$ deviation of the small filled blue triangles
(red circles). Black ``X'' with error bars, identical to the red symbols in
Figure \ref{fig:colorvar}, show the mean and 1$\sigma$ deviation of the case
that a global {\it constant} background is subtracted. Large empty blue
triangles (red circles) are shifted along the x-axis for clarity. The typical
color uncertainty of each clump is not shown, but can be inferred from the
uncertainty of Figure \ref{fig:colorvar}. Solid and dashed green lines 
show the mean and standard deviation of color variation of interclump
pixels. The best-fit slope of each color gradient is also given in the figure. \\
\label{fig:bkgsub_color}}
\vspace{-0.2cm}
\end{figure}

\begin{figure}[htbp]
\center{\includegraphics[scale=0.4, angle=0]{./fig9.ps}}
\caption[]{Radial variation of physical properties of clumps as a function of
galactocentric distance, under different schemes of diffuse background 
subtraction. Symbols and colors are same as those in Figure
\ref{fig:bkgsub_color}. \\
\label{fig:bkgsub_prp}}
\vspace{-0.2cm}
\end{figure}

Since the background subtraction scheme would most affect the radial variations
of color and physical properties of clumps, we investigate these variations in
another two subtraction schemes: local subtraction and no subtraction at all.
In the {\it local} subtraction scheme, for each clump, we measure the
background surface brightness in an annulet that is 0.3\arcsec (two times the
radius of our photometry aperture) away from the clump, with all areas within a
distance of 0.3\arcsec of any other clumps masked out. Then, we subtract a
corresponding background flux from the raw flux of the clump. In the {\it zero}
scheme, we do not subtract any background from the raw fluxes of the clumps. We
re-derive the rest-frame colors and physical properties of the clumps and
re-analyze their radial variations. The new results are shown in Figure
\ref{fig:bkgsub_color} for rest-frame UV/optical color and Figure
\ref{fig:bkgsub_prp} for physical properties and compared with the results of a
global {\it constant} subtraction.

In both new subtraction schemes, the trend of the radial color variation is
still present, although being flattened slightly: the difference between the
average color of clumps at ${\rm d<0.1}$ and ${\rm d>0.5}$ becomes from 0.8 mag
in the {\it constant} subtraction (black) to 0.5 mag in the {\it local} (blue)
or the {\it zero} subtraction (red). As discussed above, the constant
background subtraction scheme may over-subtract light, especially red light,
from clumps in the outskirts, resulting in a false bluer color for them.
However, in the other side, the zero background subtraction leaves the
background light in red bands to these outskirts clumps, which would result in
a false redder color for them. The real trend of the color radial variation
should be the one between the constant subtraction and zero subtraction, most
likely the local subtraction. We also note that the color of clumps in ${\rm
d<0.1}$ is not significantly changed in our zero subtraction scheme, implying
that background emission is negligible for central clumps. Overall, the change
of the color variation under different subtraction schemes is no statistically
significant. Therefore, we conclude that the observed trend of color variation
is intrinsic for clumps.

As we discussed in Sec. \ref{prpvar:colorvar}, the host ``disks'' exhibit
a mild color gradient. It is important to examine whether the radial variation
of clumps under different background subtraction is still significant steeper
than that of interclump pixels. Since the background subtraction does not
affect the gradient of interclump pixels (as discussed before, we exclude all
pixels close to identified clumps), we simply overplot their color gradient
obtained in Sec. \ref{prpvar:colorvar} in Figure \ref{fig:bkgsub_color} (green)
and compare its slope with that of clumps under various background subtraction
schemes. The slope of clumps in the {\it zero} background subtraction is only
steeper than that of interclump pixels at 2.4$\sigma$ level. However, as
discussed above, this trend, suffering from under-subtraction of diffuse
background, can only be treated as the lower limit of the color gradient of
clumps. The most likely color gradient of clumps, i.e., the one with {\it local}
background subtraction, is 3.7$\sigma$ steeper than that of interclump pixels.
Thus, we conclude that the subtraction schemes of the diffuse background would
not change our previous conclusion that the color gradient of ``disks'' is
significantly flatter/weaker than, and hence only contribute to a small part
of, that of clumps.

The radial variations for physical properties under different subtraction
schemes are shown in Figure \ref{fig:bkgsub_prp}. The situation here is similar
to that of Figure \ref{fig:bkgsub_color}: all trends are still present, but
with strength reduced. The trend of SSFR is reduced the most and only shows a
marginal signal that clumps at ${\rm d>0.5}$ are only a 1.5--2 times higher
than those at ${\rm d<0.1}$. The average age of clumps at ${\rm d>0.5}$
increases from 100 Myr to 300 Myr so that the age difference between ${\rm
d<0.1}$ and ${\rm d<0.1}$ reduces from $>$500 Myr of Figure \ref{fig:prpvar} to
$\sim$300 Myr. The trend of dust extinction is almost unchanged, implying that
the E(B-V) measurement is statistically robust against background subtraction.
The trend of the stellar surface density is also flattened, but still
significant, even compared with the trend in Figure \ref{fig:prpvar}. The
stellar surface densities of central clumps are now on average 10 times higher
than that of clumps in outskirts. 

Similar as for the color, the changes of above trends on physical properties
are largely caused by the change of the derived properties of clumps in
outskirts. On the other side, properties of central clumps are less (almost
not) affected by the employed background subtraction schemes. Compared to the
extreme case of over-subtraction due to the assumption of constant background,
the zero subtraction is another extreme case for clumps in outskirts. The real
diffuse background should be somewhere between these two cases. If the correct
background is subtracted for these clumps, we would expect the radial
variations of color and physical properties be stronger than those under the
zero subtraction, but weaker than those under the constant subtraction.
Overall, we can still conclude that central clumps are redder, older, denser,
more obscured, and less active in forming stars than clumps in outskirts.

\subsection{Possible Metallicity Variation}
\label{prpvar:metallicity}

Our above analyses are based on our SED-fitting assumption that all clumps have
solar metallicity. However, if there is a radial trend in metallicity (rich
center and poor outskirts) already in place in the underlying disk, clumps that
form closer to the center would tend to be redder, and vice versa.  The radial
variations due to the underlying metallicity gradient might mimic the color
trends expected from the clump migration scenario.

The spatial variation in metallicity is indeed observed in clumpy SFGs at
z$\sim$2. \citet{genzel08} found that three out of five their SINS galaxies
exhibits a radial gradient of the [N II]/H$\alpha$ ratio, which implies a
$\sim$20\% higher oxygen metallicity in the central region than in the outer
disks. Furthermore, \citet{genzel11} found clump to clump and center to outer
variation in the [N II]/H$\alpha$ ratio in another three SFGs at z$\sim$2.
These findings are broadly consistent with the inside--out growth mode
predicted by semi-analytic models \citep[e.g.,][]{somerville08}. However, we
note that the metallicity variation in their works is modest though.  For
example, the average metallicity among the three galaxies increases from
$\sim$0.5 $Z_\odot$ in outer disks to $\sim$0.8 $Z_\odot$ in central regions.

Even if the above metallicity variation exists in our galaxies, its
contribution to our observed color variation would be small. The rest-frame U-V
color of a constant star-forming template with age of 1 Gyr changes only 0.2
mag from $Z=Z_\odot$ to 0.2$Z_\odot$. However, in Figure
\ref{fig:bkgsub_color}, the radial color variation is $\gtrsim$0.6 mag,
requiring other variations to explain it. We acknowledge the existence of
metallicity variation, but still conclude that assuming a constant metallicity
would not significantly affect our results on the radial variations of the
SED-fitting derived properties. Moreover, the modest variation cannot be well
constrained by SED-fitting, because the usually employed discrete metallicity
distribution: sub-solar, solar, and super-solar, is too broad to describe the
variation. 

\section{Discussion}
\label{discuss}

\subsection{Comparisons with Other Studies}
\label{disc:other}

We compare our measurements of the properties of clumps with those of other
studies. Due to the differences on a few key ingredients, e.g., sample
selection, photometric apertures, background subtraction schemes, and models
used to derive physical properties, only order-of-magnitude comparisons can be
made. We focus on their stellar masses, SFRs and contributions to host
galaxies. We are not able to compare clump sizes with other studies, as we fix
a constant photometric apertures for all clumps.  In order to compare
stellar masses and SFRs that are derived with different IMFs, we apply the
relation of \citet{salimbeni09} to convert results with the Chabrier IMF
\citet{chabrier03} to that with the Salpeter IMF: ${\rm log(M_{Salpeter}) =
log(M_{Chabrier}) + 0.24}$. Since the normalization of SFR is determined by
stellar mass in SED-fitting, we also scale the SED-fitting derived SFRs by a
relation deduced from the above one: ${\rm SFR_{Salpeter} = 1.74 \times
SFR_{Chabrier}}$.

Since three galaxies in our sample (Galaxy 21739, 22284, and 27101) have
been studied by \citet{elmegreen05}, with ID 3465+, 3483, and 6462+ in their
paper, we first compare the derived properties of these galaxies in the two
studies in details, and then compare clump properties in sample wise with other
studies. Table \ref{tb:comp} shows the first comparison. As discussed in Sec.
\ref{data:sample}, \citet{elmegreen05} used ACS i-band to select clumpy
galaxies and identified more clumps in each galaxy than we do. In order to
derive physical parameters, including redshifts, of clumps as well as galaxies,
\citet{elmegreen05} compare ACS color pairs of galaxies and clumps to that of
stellar population synthesis models with exponentially decline SFH. Although
the derived properties of clumps are different between the two studies and we
are limited to a very small sample, we do not find systematical difference
between the two studies for the following properties: average age of clumps,
average stellar mass of clumps, stellar mass of host galaxies, and the fraction
of mass in clumps. 

The two parameters with obvious offsets between the two studies are the age of
disks (called interclump age in \citet{elmegreen05}) and average SFR of clumps.
The interclump age in \citet{elmegreen05} is $\sim$2 Gyr, while the disk age in
our study is 0.3--0.5 Gyr. The difference could be due to the assumption of
\citet{elmegreen05} that interclump star formation began at z=6. It could also
be caused by the degeneracy between age ($t$) and the characteristic decay 
timescale ($\tau$) in the SED-fitting of either of the two studies. However, the
difference in the disk age would not change the results in both studies. 
\citet{elmegreen05} found that clumps are bluer than interclump regions,
consistent with our conclusion that clumps are spots with enhanced SSFR. The
systematical offset in the other parameter, the average SFR of clumps, is
likely caused by the use of different SFR indicators in the two studies.
\citet{elmegreen05} used the instantaneous SFR at a given time $t$, while we
average the SFR over the last 100 Myr according to the SFH of the best-fit
model. We argue that such average is necessary if one wants to compare the SFR
derived from SED-fitting to that empirically derived the rest-frame UV
continuum, as the lifetime of O and B type stars is about 100 Myr. Since
\citet{elmegreen05} used the exponentially declining model, such average will
elevate the SFR of clumps from the instantaneous value of $\sim$1 ${\rm M_\odot
yr^{-1}}$ to several ${\rm M_\odot yr^{-1}}$, consistent with our measurements.
Overall, we conclude that the measurements of clump properties and, more
importantly, the interpretation of the properties in the two studies are
broadly consistent.

Now, we compare clump properties statistically in sample wise with other
studies. The stellar masses of our clumps agree very well with those of
\citet{fs11b}, but are slightly larger those that of
\citet{elmegreen05,elmegreen09a}.  The clumps in the NIC2 sample of
\citet{fs11b} span the stellar mass range of ${\rm 10^{8} M_\odot}$ to ${\rm
10^{10} M_\odot}$, with a median mass of about ${\rm 3 \times 10^{9} M_\odot}$.
\citet{elmegreen05} found a typical mass of ${\rm 6 \times 10^{8} M_\odot}$ for
clumps in 10 HUDF galaxies.  However, a revisit of HUDF by \citet{elmegreen09a}
with NICMOS images shows that clumps are typically in range of ${\rm 10^{7}
M_\odot}$ to ${\rm 10^{9} M_\odot}$. It is possible that the difference of
stellar mass is resulted from sample selections. In our sample, we only choose
galaxies with spectroscopic observations, which would bias our sample toward
UV/optical luminous (and hence massive) end, while \citet{elmegreen09a} did not
restrict their sample to spec-z and hence were able to detect clumps in fainter
galaxies. As discussed in \citet{genzel11}, the Toomre mass of clumps is
proportional to the mass of disks. Clumps detected in our possibly biased
massive samples would therefore have higher clump masses. 

The SFRs of clumps in our study are broadly consistent with those of
\citet{genzel11}, who measured the SFRs of clumps through extinction corrected
H$\alpha$ luminosity. Their SFRs (after being converted to a Salpeter IMF)
run from a few ${\rm M_\odot yr^{-1}}$ to $\sim$70 ${\rm M_\odot yr^{-1}}$,
with a median (mean) of 24 (28) ${\rm M_\odot yr^{-1}}$. In our sample, the
SFRs of majority clumps cover a similarly wider range, from less than 1 ${\rm
M_\odot yr^{-1}}$ to $\sim$70 ${\rm M_\odot yr^{-1}}$, with a median (mean) of
$\sim$10 ($\sim$24) ${\rm M_\odot yr^{-1}}$. The consistency of SFRs in the two
studies is encouraging, because \citet{genzel11} and we have used two different
physical mechanisms (nebular line emission and stellar color) to measure SFRs.
\citet{fs11b} only measured SFR for galaxies as a whole. If we scale the SFRs
of their host galaxies by the clump fractional contribution to H-band light in
their study, typically 5\%, the derived typical SFR of clumps is only a few
${\rm M_\odot yr^{-1}}$, smaller than our median value.  However, as we
discussed in Sec.  \ref{prp:cvg}, as regions with enhanced SSFRs in ``disks'',
clumps contribute more on SFR than on stellar mass (and hence H-band light) to
host galaxies. Therefore, the actual SFRs in clumps of \citet{fs11b} should be
higher than the above scaled value, moving them closer to our measurement.  In
fact, one galaxy (BX 482) in \citet{fs11b} has H$\alpha$ measurement for each
of its clumps. We calculate the H$\alpha$ derived SFRs for each of its clumps.
Since \citet{fs11b} did not report extinctions for individual clumps, we
assume that the dust extinction of each clump is equal to the global extinction
of BX 482, ${\rm A_V = 0.8}$, measured through SED-fitting. We find that SFR
varies from 5 ${\rm M_\odot yr^{-1}}$ to 46 ${\rm M_\odot yr^{-1}}$ for clumps
in BX 482, with a median (mean) of 9 (14) ${\rm M_\odot yr^{-1}}$, broadly
consistent with our clump SFRs. Finally, we note that H$\alpha$ is a measure of
instantaneous SFR more than of the averaged SFR over the last 100 Myr, which we
derive from our SED-fitting. The difference of the two SFR indicators may
induce a bias when we compare the two types of SFRs. However, given the large
uncertainties of measuring SFRs in both ways as well as the large scatter of
SFRs of clumps, such a bias would not significantly affect our conclusions.
SFRs measured through both H$\alpha$ and SED-fitting on larger samples are
needed for carrying out more precise comparisons to better understand the star
formation process in clumps.

In terms of fractional contribution of light to host galaxies, we compare our
measurement on rest-frame UV/optical light to that on the observed emission of
\citet{fs11b}. At z$\sim$2, the rest-frame U and V band emissions are
redshifted to close to the $H_{160}$ and $i_{814}$ bands in the observer's
frame, validating the comparisons between the two papers. In our study, the
typical individual clump contribution to the UV/Optical luminosity is $\sim$5\%
(spreading over a wide range of 1\% to 10\%) and the total contribution of
clumps in a galaxy is $\sim$20\%. We also find no significant difference
between the clump contributions to U-band and V-band light. These results are
quite similar to what \citet{fs11b} found with the $H_{160}$ fluxes of their
NIC2 clumps as well as $i_{814}$ emission in one of their galaxy. The
contribution on ACS $i_{775}$ band emission by clumps in \citet{elmegreen05}
and \citet{elmegreen09a} is typical 2\%, slightly smaller than our average
value.  However, given that galaxies in \citet{elmegreen05} and
\citet{elmegreen09a} usually contain 5--10 clumps, about two times more than our
galaxies, the total contribution of clumps to host galaxies is about 25\%,
close to our value. \citet{wuyts12} found that the fractional contribution
of clumps to the total light of their hosts has a mild dependence on the
waveband used for identifying clumps, decreasing with the increasing of
wavelength. They also found that for a giving identification band, the
fractional contribution of clumps slightly decreases with wavelength. In their
z$\sim$2 sample, the fractional contribution of clumps to the total rest-frame
U (V) light of their hosts is, on average for detections in 2800\AA, U, and V
bands, 20\% (17\%), close to our values. However, we note that besides using a
different sample selection criterion (mass-complete), \citet{wuyts12} have two
other approaches that are different from ours: excluding the central bulge of a
galaxy as a clump and not subtracting diffuse background of hosts from clump
light, each of which brings an effect to the fractional contribution opposite
to the other. These results again highlight the fact that the contribution of
clumps to the total light of their host galaxies is small, and the light
distributions of galaxies are still dominated by diffuse components.

\subsection{Formation of Clumps}
\label{disc:formation}

In the commonly assumed framework that giant clumps are formed through
gravitational instability of turbulent gas-rich Toomre-unstable disks
\citep[e.g.,][]{noguchi99,immeli04a,immeli04b,bournaud07,bournaud08,elmegreen08,genzel08,dekel09,ceverino10,genzel11,fs11b},
clumps have a characteristic scale and mass, namely the Toomre length and the
Toomre mass. They represent the largest and fastest growing unstable mode
that is not stabilized by rotation. Since we use a fix photometric aperture
for all clumps, we lose the size information of clumps. However, we can still
compare our clump masses with the Toomre mass predicted by the in-situ
fragmentation scenario. We use Equation (5) in \citet{genzel11} to calculate
the Toomre mass of disks: 
\begin{equation} 
  M_{Toomre} \approx 5 \times 10^9 \left(\frac{f_{young}}{0.4}\right)^2 \left(\frac{M_{disk}}{10^{11}M_\odot}\right) M_\odot, 
\label{eq:toomre}
\end{equation} 
where $f_{young}$ is the mass fraction of component of stars, and
$M_{disk}$ is the total mass of disk, which is close to the baryonic (gas + star)
mass in the central $\sim$10 kpc of galaxies. The maximum clump mass that can
be formed in a uniformly rotationally supported gas disk is actually first
derived by \citet{escala08}:
\begin{equation} 
  M_{cl}^{max} = 3 \times 10^7 M_\odot \left(\frac{\eta}{0.2}\right)^2 \left(\frac{M_{gas}}{10^{9}M_\odot}\right), 
\label{eq:clmax}
\end{equation} 
where $\eta = M_{gas}/M_{tot}$ is the ratio of the gas mass to the total mass
enclosed with a radius $R$. The maximum clump masses estimated by the 
two equations are in 
agreement within a factor of two. \footnote{We note that the maximum clump
mass is not the turbulent Jeans mass, which is often incorrectly interpreted in
some of previous studies. The Jeans mass (and Jeans length), beyond which the
disk cannot be stabilized purely by thermal pressure, is actually the smallest
unstable mode that can be formed in a disk. We refer readers to
\citet{escala08} for detailed discussions.}

To apply Equation \ref{eq:toomre} to our galaxies, we assume a
gas--to--baryonic mass fraction of 0.5, close to the median value of the
fraction observed by several authors
\citep[e.g.,][]{erb06,genzel08,tacconi08,tacconi10,fs09,daddi10}. Therefore, we
have $f_{young}=0.5$ and $M_{disk}$ is two times the stellar masses of the
disks that are measured through SED-fitting. We compute the Toomre mass for
each galaxy and compare the masses of clumps in the galaxy to the Toomre mass.
The ratio between clump mass and Toomre mass spans a wide range, from 0.05
to 3.5, with a median of 0.3. This result is encouraging, as it demonstrates
that our clump masses are broadly consistent with the characteristic mass
predicted by the scenario of disk instability. The statistically smaller
masses of clumps are not contradictory with the prediction, as the Toomre mass
is the maximum unstable mass. Moreover, Toomre mass is proportional to the
third power of gas density and inversely proportional to the forth power of
angular rotation speed \citep[see Equation (2) of ][]{escala08}. Since both gas
density and angular rotation speed are functions of radius, the Toomre mass
also varies with galactocentric distance.  In a disk with flat rotation curve
and gas density decreasing with radius, the Toomre mass decreases with radius.
Our clump mass distribution is also consistent with this argument, as less
dense (and hence less massive with given photometric aperture) clumps are found
at large galactocentric distances.  Overall, the clump masses in our sample are
broadly consistent with the prediction of the scenario of gravitational
instability.

The in-situ fragmentation due to gravitational instability requires a gas-rich,
turbulent and marginally unstable ($Q\sim$1) disk as the birthplace of clumps.
The existence of such disks cannot be directly inferred from our multi-band
images; it has to be confirmed through the kinematics of star, gas or ISM.
\citet{fs09} presented the spatially resolved gas kinematics of 62 star-forming
galaxies at z$\sim$1--3, measured through H$\alpha$ and [N II] emission lines
observed by the Spectroscopic Imaging survey in the Near-infrared with SINFONI
(SINS). They found that about one-third of galaxies in their sample are
rotation-dominated yet turbulent disks, another one-third are compact and
velocity dispersion-dominated objects, and the remaining one-third are
interacting or merging systems. The also found that the fraction of
rotation-dominated systems increases toward the massive end of the sample.
Since almost all our galaxies have stellar mass larger than ${\rm 10^{10}
M_\odot}$ and lie on the massive end of the mass spectrum of SINS sample, we
expect that the fraction of rotation-dominated turbulent disks in our sample is
higher than 40\%. 

In fact, one galaxy in our sample, 27101, was observed by \citet{bournaud08}
through H$\alpha$ field spectroscopy using SINFONI on VLT UT4. They found a
large-scale velocity gradient throughout the system, with large local kinematic
disturbances. They also found a disk-like radial metallicity gradient in the
galaxy. These findings can be most likely explained by the scenario of internal
disk fragmentation, despite the complex asymmetrical merger-like morphology.
However, another galaxy in our sample, 24919, shows an obvious
interaction/merger signature. A long (tidal) tail is curving from its lower
left part all the way to its upper middle part, as can be seen in its z-band
image in Figure \ref{fig:mosaic}, suggesting an ongoing interaction/merger,
which might be responsible for the formation of clumps. To understand to which
extension the in-situ fragmentation scenario is valid to explain the formation
of giant clumps requires a large survey of kinematics of clumpy galaxies, in
addition to their multi-band images.

Another possible interpretation of the clumpy features in our sample
galaxies is that these clumps do not actually represent any physical entities,
but simply correspond to locations with lower line-of-sight dust obscuration in
the host galaxies. This alternative arises from the fact that these clumps are
bright in the rest-frame UV images. If this interpretation is true, the
properties (e.g., age and SSFR) and observed radial variations of clumps would
plausibly reflect those of the underlying galaxy population. Based on our
previous results, however, we argue that this alternative interpretation is
unlikely true and that clumps are physical entities with properties and
formations differing from those of their host ``disks''. For clumps in the
outskirts (e.g., $r > 0.1 r_{Kron}$) of galaxies, it is true that their dust
obscuration is lower than that of interclump regions at the same galactocentric
distance (the bottom left panel of Figure \ref{fig:prpvar}. However, these
clumps also have systematically younger ages than their nearby interclump
regions. The very young age ($\lesssim$100 Myr) of these clumps indicates that
they are newly formed, possibly due to the instability induced by the cold
accretion, which preferentially occurs in the outskirts of galaxies, in a
relatively older (and hence stable) ``disks''. For central clumps, if they
represented the underlying ``disk'' stellar populations but had lower dust
extinction, their rest-frame colors should be {\it bluer} than their
surrounding areas. However, this expectation is contradictory to our previous
result, namely the rest-frame UV color of clumps is redder than that of central
interclump regions (Figure \ref{fig:colorvar}). In fact, the bottom left panel
of Figure \ref{fig:prpvar} shows that the dust extinction of central clumps is
comparable to (or even higher than) that of central ``disk'' regions, if our
SED-fitting technique does not significantly suffer from the age--extinction
degeneracy. Overall, it is unlikely that the appearance of the clump features
is simply due to lower line-of-sight extinction. As a result, clumps should
have origins distinctive from that of ``disk'' stellar populations.

\subsection{Fate of Clumps} 
\label{disc:formation}

There are two possible scenarios commonly proposed to explain the fate for
giant clumps in z$\sim$2 SFGs: they would (1) migrate toward the gravitational
centers of their host galaxies due to interactions and dynamical friction
against the surrounding disks and eventually coalesce into a young bulge as the
progenitor of today's bulges or (2) be rapidly disrupted by stellar feedback,
supernova feedback, or tidal torques during (or even before the beginning of)
their migration toward centers. \citet{dekel09}  made a few predictions that
observations can test for the possible fate of clumps. If the migration
scenario is true and clumps survive for a migration timescale of $\sim$0.5 Gyr, 
giant clumps would (1) have an age spread of $\sim$0.5 Gyr; (2) be gas rich and
forming stars at a high rate that is similar to the preceding few hundred Myr;
and (3) have a radial age variation in the sense that clumps at large disk
radii are younger than $\sim$0.5 Gyr, while those at smaller radii are older.
In contrast, if clumps are rapidly disrupted, they would (1) have smaller age
spread, $\sim$100 Myr and (2) have no obvious age gradient with galactocentric
radius. 

Our findings on the properties of giant clumps are reasonably consistent with
the prediction of the inward migration scenario. The top right panel of Figure
\ref{fig:prpvar} shows that the age distribution of our clumps
spans a wide range from $<$0.1 Gyr to a few Gyr. The age spread is comparable
to the prediction of $\sim$0.5 Gyr of the migration scenario, but significantly
larger than the prediction ($\sim$ 100 Myr) of the disruption scenario. If no
background subtraction is applied to our clumps, the age spread (see in Figure
\ref{fig:bkgsub_prp}) is even closer to the prediction of the migration
scenario. The radial age variation of our clumps is also strongly in favor of
the migration scenario. Recently, \citet{ceverino12} discussed the
internal support of the in-situ giant clumps in gravitationally unstable disks
at high redshift, using both an analytic model and high-resolution hydro
adaptive mesh refinement simulations. They predicted a steep age gradient of
clumps throughout their host disk due to the formation of giant clumps in the
outer parts of the disk and their inward migration to form a bulge in the disk
center. Our results agree very well with their predictions: clumps at $d>0.5$
have the mean age of $\sim$100 Myr, while those at $d<0.1$ $\sim$ 700 Myr.
\citet{ceverino12} even predicted that the age gradient of clumps is steeper
than that of interclump stars. Such prediction is similar to our results in
Figure \ref{fig:prpvar} and hence strengthens our argument on the inward
migration scenario. Another piece of evidence, interesting but largely
uncertain, is coming from the SFH of clumps. As described in Sec.
\ref{detection}, we fit each clump with three types of SFHs: exponentially
declining, exponentially increasing and constant. We then choose the most
likely SFH based on the reduced $\chi^2$ of the best-fit of each SFH. About
70\%--80\% of clumps are thus classified as having constant SFH, agreed with
the above prediction of the migration scenario of \citet{dekel09}. However, we
note that the SFH derived from SED-fitting is approximate and severely model
dependent and can only be used as a loose constraint on predictions. Overall,
we conclude that the age spread and radial variation indicate that these clumps
might eventually migrate into the centers of their host galaxies. 

It is also possible, however, that not all clumps can survive long enough to
migrate into the gravitational centers. Some of them might still be disrupted,
possibly by the stellar feedback (while the effect of supernovae feedback seems
unimportant (see \citet{dekel09})). \citet{genzel11} observed strong outflows
in their clumps, with a rate as large as or even larger than SFRs.  They also
estimated the gas expulsion time due to outflows, which ranges from 170 to 1600
Myr from clump to clump. A hint of the disruption of clumps can also be
inferred from the radial age variation of clumps (the top left panel of Figure
\ref{fig:prpvar} and Figure \ref{fig:bkgsub_prp}). In this figure, we find no
clumps with age $\lesssim$100 Myr at small galactocentric radius ($d<0.1$). The
reason of the dearth of young central clumps could be either that young clumps
are preferentially formed at large radii or that young clumps at small radii
are rapidly disrupted due to the somehow stronger outflows or interactions. If
the latter is true, the disruption timescale (or lifetime) of clumps should
have a relation with the their densities, because due to their shallower
potential wells and less concentrated structures, low density clumps are easier
to be disrupted by either outflows or tidal torques than high density clumps.

\begin{figure}[htbp]
\center{\includegraphics[scale=0.8, angle=0]{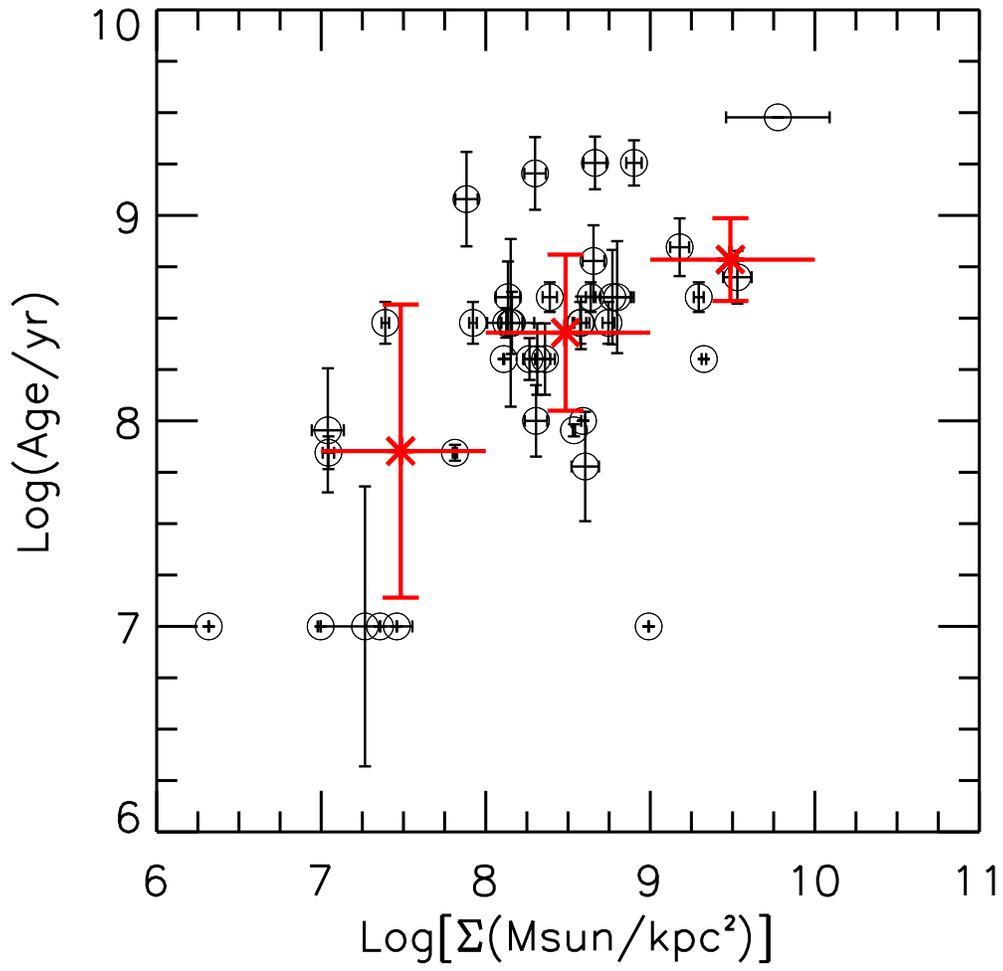}}
\caption[]{Age--stellar surface density relation of clumps.  Symbols and colors
are same as those in Figure \ref{fig:colorvar}. A constant background has been
subtracted for clumps in this figure. \\
\label{fig:ageden}}
\vspace{-0.2cm}
\end{figure}

Figure \ref{fig:ageden} shows that the age of clumps increases
with the stellar surface densities of clumps. The upper envelope of the
relation provides a rough estimate on the lifetime for clumps with different
stellar surface densities. For example, no clumps with density of ${\rm 10^{7}
M_\odot/kpc^2}$ are older than 100 Myr, suggesting that the lifetime of this
type of clumps is $\sim$100 Myr. While for clumps that are one order of
magnitude denser, their lifetime could be up to 1 Gyr. However, we
note that we cannot rule out an alternative scenario that all clumps are
formed at large radii and become old and concentrated when they migrate toward
galactic centers.

\subsection{Ongoing Bulge Formation?}
\label{disc:bulge}

If the inward migration scenario is true, one would expect to find bulges or at
least ongoing forming bulges in a fraction of z$\sim$2 SFGs. Clumps are
observed within a wide redshift range, from z$\sim$4 \citep{elmegreen07} to
z$\sim$2 \citep[e.g.,][]{genzel08,genzel11,fs11b} or to even lower redshift. Moreover, very young clumps
are observed at z$\sim$2 in our sample or other studies \citep[e.g.,][]{fs11b}.
These observations show that the formation of clumps is a continuous process
that at least lasts over the cosmic time of $\sim$1.7 Gyr from z$\sim$4 to
z$\sim$2. If we assume a constant clump formation rate over the cosmic time and
a bulge formation (due to the coalescent of clumps) timescale of 1 Gyr, a few
times the migration timescale of clumps, the fraction of SFGs that contain a
young bulge at z$\sim$2 should be (1.7-1)/1.7$\times$100\%$\sim$40\%. 
If the clump formation rate increases with redshift, the bulge
fraction would be lower, as the majority of newly formed clumps have
insufficient time to coalesce into bulges, and vice versa.

We roughly estimate the bulge fraction in our sample through the following
ways: 

\begin{enumerate}

\item {\it Morphology and color}: Bulges are believed to have spheroid-like
morphology and be redder than other components if they are older. They also
tend to reside in the gravitational centers of galaxies. By simply looking at
the mosaics of our sample in Figure \ref{fig:mosaic}, we identify five galaxies
(21852, 22284, 24033, 24684, 27101) that contain such a component that
satisfies the above conditions. This bulge fraction (50\%) is broadly
consistent with what we estimated above based on the bulge formation timescale
(40\%). We note that although morphology and color are a relative easy way to
identify bulges, they suffer from the age--dust degeneracy as well as the
problem of subjectivity. A more accurate bulge identification requires other
pieces of information, such as the kinematic information of bulges.

\item {\it Stellar mass}: As we discuss in Sec. \ref{disc:formation}, the
Toomre mass is the maximum unstable mass that the Jeans-unstable mode not
stabilized by rotation can generate. If the mass of a clump is far larger than
the Toomre mass in its host disk, the clump may not be a direct result of disk
fragmentation. Instead, it could be a coalesced result of a few clumps. In our
study, a clump is considered a bulge if its stellar mass is three times larger than
the Toomre mass. In our sample, four galaxies (23013, 24033, 24684, 27101)
contain such super-Toomre mass clumps. Moreover, these super-Toomre mass clumps
are all close to the H-band light centers --- an approximation of the
gravitational centers --- of their host galaxies. This sub-sample is largely
overlapped with the sub-sample that we identify through morphology and color,
with 3/4 of the former being in the latter. The bulge fraction of 40\% measured
in this way also agrees very well with our prediction. However, we note that
using stellar and Toomre mass to set constraint on the bulge fraction
critically relies on the accuracy and interpretation of the Toomre mass.
Therefore, it can only be treated as a rough estimate.

\item {\it X-ray detection}: It is now widely accepted that super-massive black
holes (SMBHs) are residing in the center of massive galaxies and co-evolve with
bulges. The masses of SMBHs are observed to correlate strongly with both the
fourth power of the velocity dispersion of bulges and the first power of the
bulge masses.  Also, recent theoretical models predict that the internal
violent processes, such as clump--clump interaction, tidal force, etc., would
feed gas to the center of galaxies to form bulges as well as SMBHs
\citet{bournaud11}. If bulges have been formed in our sample, we would expect
some of our galaxies to be detected as AGN in X-ray due to the energetic
feedback released by their accompanying SMBHs.  Indeed, four galaxies (23013,
24684, 24919, 27101) in our sample have significant detections in the Chandra
4Ms imaging of CDFS\footnote{http://cxc.harvard.edu/cda/Contrib/CDFS.html},
with three detected in both the soft and hard bands and one (27101) only in the
soft band. At such high redshift (z$\sim$2), all these X-ray sources have high
luminosities that can only be generated via AGN (Lx$>10^{42}$ ${\rm erg
s^{-1}}$). This finding is encouraging, since this X-ray detected sub-sample is
prominently overlapped with the super-Toomre mass sample that we just discussed
above, with 3/4 of the former is found in the latter. The only exception,
Galaxy 24919 shows an extraordinary red lane in its z-H color map.  Regardless
of the component of the lane, i.e., old stars or dust, its existence indicates
that the galaxy is undergoing a dramatic violent process, possibly merger, a
common way to form bulges. Including that, the bulge fraction inferred from
X-ray detection is well agreed with our prediction of 40\% as well as that
induced from Toomre mass.

\item {\it Age}: Bulges should contain old stellar populations with ages older
than a few times of the migration timescale of clumps ($\sim$0.5 Gyr).  In our
sample, only two galaxies (24686 and 27101) contain clumps with age older than
1 Gyr. Moreover, these clumps are very close to the centers of host galaxies.
These findings indicate that these clumps could be bulges rather than newly
formed clumps. The two galaxies also satisfy all above three criteria.
Therefore, age provides the most restricted constraint on the bulge fraction
and yields the smallest fraction (20\%).  However, we note that the age
measured through SED-fitting is actually a light-weighted average age for both
old and young population. In our clumps, since their SFRs have not been fully
quenched, the new formed population would drag the measured age toward the
young side. Therefore, some other clumps with age close but less than 1 Gyr may
also contain populations as old as 1 Gyr and hence be candidates of
proto-bulges.

\end{enumerate}

Overall, our estimations yields a bulge fraction in our sample of z$\sim$2
clumpy galaxies from 50\% through morphology and color to 20\% through age. The
most likely fraction of 40\%, obtained by the prominent overlapping of
super-Toomre mass and X-ray detection, is very close to our prediction based on
the migration timescale of clumps. This result suggests that bulges have been
likely formed through the coalescence of giant clumps in our sample of z$\sim$2
clumpy galaxies and this process is still going on.

However, one should be cautionary when generalizing the conclusion to the
general bulge formation of SFGs at z$\sim$2. First, our sample is limited to
only ten galaxies selected to be particularly clumpy and is biased toward UV
bright, blue, and large galaxies, as discussed in Sec. \ref{data:sample}.
Second, our work cannot fully rule out other processes as the mechanisms of
general bulge formation at z$\sim$2. 
\citet{genzel08,genzel11} studied the possible rapid disruption of clumps via
vigorous outflows and the kinematic signatures of inward gas streaming motions
in clumpy disks. These signatures suggest that the gas inflows that are fed to
the centers of galaxies through internal violent processes may play an
important role in building bulges along with clump migration, as highlighted in
some of the most recent numerical simulations \citep[e.g.,][]{bournaud11}.
Furthermore, given the fact that some objects in our sample may be merging
systems, e.g., Galaxy 24919, the role of merger on building bulges cannot be
fully excluded. It is very likely that more than one process could be
responsible for early bulge building at z$\sim$2. To understand the
contribution of each possible mechanism requires further investigations with
larger samples and newer observations.

\section{Summary and Conclusion}
\label{summary}

In this paper, we study the properties of kpc-scale clumps in z$\sim$2
SFGs through broad band multi-wavelength photometry. We identify clumps through
a hybrid of auto-detection and visual inspection from ten galaxies with spec-z
between 1.5 and 2.5 in the HUDF, where the recently available deep HST/WFC3
images, together with HST/ACS images, enable us to resolve into kpc-scale at
z$\sim$2 and detect clumps toward the faint end. Using the spatially resolved
seven-band (BVizYJH) photometry, we measure physical properties of clumps
through SED-fitting and allow the fitting to choose the best-fit parameters
through three types of SFHs: exponentially decreasing, exponentially
increasing, and constant. We also measure the properties of ``disks'', namely
the diffuse components of the host galaxies. The main results of our paper are
summarized as follows:

\begin{enumerate}

\item The number of clumps in our galaxies runs from 2 to 5, with a median of
4. The total number of 40 clumps enables us to study the physical properties of
clumps with sufficient number statistics. 

\item Most of our clumps have blue rest-frame UV--optical colors that are
similar to the colors of their surrounding ``disks'' or their host galaxies as
a whole. Only few clumps are as red as quiescent galaxies at z$\sim$2. The CMD
of clumps indicates that they are still actively forming stars.

\item The SFR--stellar mass relation of clumps and ``disks'' have almost the same
slopes, but that of clumps has higher normalization, converting to higher SSFR.
However, the SFR--stellar mass relation of the host galaxies as a whole is
still dominated by ``disk'', as both SFR and stellar mass of host galaxies are
largely contributed by ``disks''. This shows that clumps are regions with
enhanced SSFR in ``disks''. 

\item An individual clump typically contributes a few percent of the rest-frame
UV/optical luminosity and stellar mass to its host galaxy. Together, all clumps
in one galaxy typically contribute 20\% of the luminosity and stellar mass of
the host galaxy. The contribution of clumps on SFR is higher, individually
about 10\% and together 50\% of the host galaxies, consistent with the fact
that clumps are regions with enhanced SSFR.

\item Clumps differ from their surrounding area in terms of age and stellar
surface density. On average, clumps are younger by 0.2 dex and denser by eight 
times than their ``disks''. There is no obvious difference between the E(B-V)
distributions of the two components. In terms of rest-frame U-V color, both
components have similar median values, but clumps spread over a broader range,
which may indicate the different SFHs or evolutionary stages of clumps. 

\item Clumps have obvious radial variations on their properties. Clumps close
to the centers of their host galaxies (the Kron radius scaled projected distance
$d<0.1$) are 0.7 mag redder in rest-frame U-V color than those in outskirts
($d>0.5$). Spatially resolved SED-fitting shows that the color trend can be
explained by the combination of radial variations of age and dust extinction.
Central clumps ($d<0.1$) are typically 600 Myr older and more extincted (E(B-V)
larger by 0.2) than those in outskirts, with the latter is typically 100 Myr
old and having E(B-V)$\sim$0.2. The central clumps are also 25 times denser
than their outskirt counterparts. However, the trend of SSFR is slightly weak,
only increasing by a factor of five from the central to outskirts. 

\item The radial variations are affected by the scheme of diffuse background
subtraction, but our conclusions are unaffected. Besides the above results with
a {\it constant} background subtraction, we also study the radial variations
under other two background subtraction schemes: {\it local} subtraction and
{\it zero} subtraction. All the trends observed in the {\it constant}
subtraction scheme are reserved in the two new cases of subtractions, but with
strength weaken. The change of the strength is largely caused by the redder
color of outskirt clumps in new subtraction schemes than in the {\it constant}
subtraction scheme. The central clumps are now only 0.5 mag redder and 300 Myr
older than outskirt clumps. The trend of E(B-V) does not significantly
change. And the trend of stellar surface density is still prominent, with the
central ones still 10 times denser than the outskirt ones.  The SSFR trend is
weaken the most, with a difference only a few times between centers and
outskirts. Arguing that the real trends would reside between the two extreme
cases of background subtraction: {\it constant} and {\it zero}, we conclude
that central clumps are redder, older, more extincted, denser, and less active
on forming stars than outskirt clumps.

\item The host ``disks'' (interclump regions) exhibit mild gradients of
rest-frame UV color, age, and dust extinction. However, these gradients are
significantly weaker/flatter than those of clumps.  The results suggest that
(1) the observed color and property gradients of clumps are not a simple
reflection of those of host "disks"; (2) the color and property gradients of
"disks" contribute to, if any, only a small part of the gradients of clumps;
and (3) the evolution of clumps is dynamically separated from those of
"disks". Thus, we claim that mechanisms such as the inward migration are
needed to explain the steeper gradient of clumps. A further test also shows
that the use of different background subtraction schemes does not significantly
change this result.

\item Our measurements of the properties of clumps are broadly consistent
with those of previous observational studies. They are also consistent with the
clump formation mechanism that clumps are formed through gravitational
instability in gas-rich turbulent disks, proposed by several theoretical work
and numerical simulations.  However, we cannot rule out the scenario that
clumps are formed through external violent processes, e.g., interactions and
mergers, as a couple of galaxies in our sample show merger signatures. 

\item The obvious radial variations of clump properties, especially the radial
age variation, are consistent with the hypothesis that clumps would migrate
toward the centers of their host galaxies, with a timescale of $\sim$0.5 Gyr,
and eventually coalesce into young bulges. However, the dearth of young clumps
in the central regions of galaxies reminds us that not all clumps are able to
survive enough to migrate into centers. We argue that the lifetime of clumps
is correlated with their stellar surface densities. Only clumps that are dense
enough can survive long enough to sink into the centers.

\item We roughly estimate whether some clumps in our sample are actually
coalesced bulges or proto-bulges through a few ways: morphology and color of
clumps, stellar mass of clumps, X-ray detection of galaxies, and age of clumps.
The bulge fraction, namely the fraction of galaxies that contain bulges, in our
sample is 20\%--50\%, depending on the way of constraint. This result is
broadly consistent with the prediction based on the migration timescale of
clumps.  We argue that the process of bulge formation is ongoing in our
z$\sim$2 sample. 

\end{enumerate}

We note that our sample only contains 10 galaxies and thus is likely subject to
small number statistics incompleteness. Also, our results apply strictly to
galaxies with relatively large UV luminosity, since our sample only includes
cases with spectroscopic redshifts. In order to obtain a robust statistical
characterization of the properties of clumps, including their dependence on
radial separation from the center of the galaxies and the fraction of clumpy
galaxies at z$\sim$2, a much larger sample covering a wider range of both
luminosity and stellar mass is needed. The ongoing CANDELS
\citep[][]{candelsoverview,candelshst} is beginning to provide deep images over
a larger sky area, $\approx 0.5$ square degree, to answer these questions.
Moreover, the deep NIR observation of CANDELS will significantly improve the
accuracy of the photometric redshift measurements at z$\sim$2, enabling us to
construct deeper samples not limited by spec-z. \citet{wuyts12} already
used part of CANDELS data to study the profiles of color, surface stellar mass
density, age, and extinction of a large sample of massive SFGs and to address
clump properties. Their approach is different from ours, as we try to identify
clumps and separate the light and properties of clumps from those of the
diffuse background of host galaxies, while they focused generally on regions
with excess surface brightness and did not subtract the diffuse background
light from clumps. Both approaches are complementary to each other and needed
to obtain a robust view on the nature and fate of clumps. In a future paper we
will report on a study similar to the one discussed in this paper that takes
advantage of the CANDELS data set.

We thank the anonymous referee for constructive comments that improve this
article. 


\newpage

\small

\begin{table*}[htbp]
\begin{minipage}[center]{\textwidth}\footnotesize
\caption{Properties of Star-forming Galaxies at z$\sim$2 \label{tb:glx}}
\begin{tabular}{cccccccccc}
\hline\hline
ID & RA & DEC & z & ${\rm M_{star}}$ & SFR & E(B-V) & U-V & M$_V$ \\
 & J2000 & J2000 & & Log(M*/M$_\odot$) & M$_\odot$/yr & & \\
\hline
  20565 &  53.1422430 & -27.7954120 & 2.016 &10.28$\pm$ 0.05 & 30.45$\pm$  6.09 &0.20$\pm$0.05 &0.64$\pm$0.00 &-21.50$\pm$  0.00\\
  21739 &  53.1485260 & -27.7969040 & 1.765 & 9.92$\pm$ 0.02 & 33.64$\pm$  6.73 &0.10$\pm$0.05 &0.23$\pm$0.00 &-21.54$\pm$  0.00\\
  21852 &  53.1492150 & -27.7788090 & 1.850 &10.35$\pm$ 0.04 & 40.35$\pm$  8.07 &0.25$\pm$0.05 &0.61$\pm$0.00 &-21.42$\pm$  0.00\\
  22284 &  53.1516190 & -27.7964320 & 1.767 &10.51$\pm$ 0.12 & 51.32$\pm$ 10.26 &0.15$\pm$0.05 &0.48$\pm$0.00 &-22.15$\pm$  0.00\\
  23013 &  53.1556430 & -27.7792950 & 1.846 &10.73$\pm$ 0.10 &162.39$\pm$ 32.48 &0.30$\pm$0.05 &0.62$\pm$0.00 &-22.51$\pm$  0.00\\
  24033 &  53.1616630 & -27.7874320 & 1.836 &10.79$\pm$ 0.15 &111.00$\pm$ 31.07 &0.35$\pm$0.06 &0.77$\pm$0.00 &-22.10$\pm$  0.00\\
  24684 &  53.1655460 & -27.7697760 & 1.552 &11.06$\pm$ 0.07 & 10.93$\pm$  6.60 &0.20$\pm$0.13 &1.36$\pm$0.00 &-21.92$\pm$  0.00\\
  24919 &  53.1668940 & -27.7987410 & 1.998 &11.02$\pm$ 0.02 &423.67$\pm$ 84.73 &0.40$\pm$0.05 &0.74$\pm$0.00 &-23.03$\pm$  0.00\\
  26067 &  53.1743180 & -27.7825190 & 1.994 &10.61$\pm$ 0.11 & 64.39$\pm$ 12.88 &0.25$\pm$0.05 &0.67$\pm$0.00 &-21.97$\pm$  0.00\\
  27101 &  53.1817060 & -27.7830010 & 1.570 &10.21$\pm$ 0.05 & 64.68$\pm$ 17.49 &0.30$\pm$0.05 &0.69$\pm$0.00 &-21.46$\pm$  0.00\\
\hline
\end{tabular}
\end{minipage}
\end{table*}

\begin{longtable}{ccccccccc}
\caption{Properties of Clumps and ``Disks''}
\label{tb:clump} \\

\hline\hline\\
  \multicolumn{1}{c}{\textbf{ID 1}} & \multicolumn{1}{c}{\textbf{ID 2}}\footnotemark[1] & \multicolumn{1}{c}{\textbf{d}} & \multicolumn{1}{c}{\textbf{${\rm \bf M_{star}}$}}\footnotemark[2] & \multicolumn{1}{c}{\textbf{SFR}} & \multicolumn{1}{c}{\textbf{E(B-V)}} & \multicolumn{1}{c}{\textbf{age}} & \multicolumn{1}{c}{\textbf{U-V}} & \multicolumn{1}{c}{\textbf{${\rm \bf M_V}$}} \\
\\
  \multicolumn{1}{c}{\textbf{Galaxy}} & \multicolumn{1}{c}{\textbf{{\small Clump}}} & \multicolumn{1}{c}{\textbf{${\rm \bf \frac{d_{proj}}{r_{kron}}}$}} & \multicolumn{1}{c}{\textbf{${\rm \bf log(M_\odot)}$}} & \multicolumn{1}{c}{\textbf{${\rm \bf M_\odot yr^{-1}}$}} & \multicolumn{1}{c}{\textbf{}} & \multicolumn{1}{c}{\textbf{Gyr}} & \multicolumn{1}{c}{\textbf{}} & \multicolumn{1}{c}{\textbf{}} \\
\\
\hline
\endfirsthead

\multicolumn{3}{c}%
{\tablename\ \thetable{} -- continued} \\
\hline\hline\\
  \multicolumn{1}{c}{\textbf{ID 1}} & \multicolumn{1}{c}{\textbf{ID 2}} & \multicolumn{1}{c}{\textbf{d}} & \multicolumn{1}{c}{\textbf{${\rm \bf M_{star}}$}} & \multicolumn{1}{c}{\textbf{SFR}} & \multicolumn{1}{c}{\textbf{E(B-V)}} & \multicolumn{1}{c}{\textbf{age}} & \multicolumn{1}{c}{\textbf{U-V}} & \multicolumn{1}{c}{\textbf{${\rm \bf M_V}$}} \\
\\
  \multicolumn{1}{c}{\textbf{Galaxy}} & \multicolumn{1}{c}{\textbf{{\small Clump}}} & \multicolumn{1}{c}{\textbf{${\rm \bf \frac{d_{proj}}{r_{kron}}}$}} & \multicolumn{1}{c}{\textbf{${\rm \bf log(M_\odot)}$}} & \multicolumn{1}{c}{\textbf{${\rm \bf M_\odot yr^{-1}}$}} & \multicolumn{1}{c}{\textbf{}} & \multicolumn{1}{c}{\textbf{Gyr}} & \multicolumn{1}{c}{\textbf{}} & \multicolumn{1}{c}{\textbf{}} \\
\\
\hline
\endhead
\endfoot

\hline \hline
\endlastfoot

\hline
  20565 & 1 &0.24 & 8.80$\pm$ 0.00 &  3.77$\pm$  0.75 &0.25$\pm$0.05 &0.20$\pm$0.00 &0.35$\pm$0.17 &-18.30$\pm$  0.14\\
 & & & 8.88$\pm$ 0.04 &  4.45$\pm$  0.89 &0.25$\pm$0.05 &0.20$\pm$0.02 &0.40$\pm$0.14 &-18.52$\pm$  0.12\\
 & & & 8.77$\pm$ 0.00 &  3.49$\pm$  0.70 &0.25$\pm$0.05 &0.20$\pm$0.00 &0.32$\pm$0.15 &-18.20$\pm$  0.13\\
  20565 & 2 &0.01 & 9.27$\pm$ 0.05 &  7.52$\pm$  2.04 &0.30$\pm$0.05 &0.30$\pm$0.04 &0.69$\pm$0.14 &-19.15$\pm$  0.12\\
 & & & 9.42$\pm$ 0.03 &  5.40$\pm$  2.01 &0.25$\pm$0.05 &0.60$\pm$0.05 &0.68$\pm$0.13 &-19.25$\pm$  0.10\\
 & & & 9.15$\pm$ 0.02 &  5.62$\pm$  1.23 &0.30$\pm$0.05 &0.30$\pm$0.02 &0.67$\pm$0.19 &-18.80$\pm$  0.15\\
  20565 & 3 &0.16 & 9.05$\pm$ 0.06 &  6.72$\pm$  2.23 &0.30$\pm$0.05 &0.20$\pm$0.03 &0.61$\pm$0.15 &-18.77$\pm$  0.12\\
 & & & 9.11$\pm$ 0.06 &  7.66$\pm$  2.55 &0.30$\pm$0.05 &0.20$\pm$0.03 &0.62$\pm$0.14 &-18.92$\pm$  0.12\\
 & & & 9.08$\pm$ 0.09 & 48.18$\pm$ 26.62 &0.30$\pm$0.05 &0.40$\pm$0.13 &0.54$\pm$0.20 &-18.62$\pm$  0.18\\
  20565 & D &---- &10.24$\pm$ 0.03 & 24.62$\pm$  4.92 &0.20$\pm$0.05 &0.90$\pm$0.07 &0.66$\pm$0.00 &-21.20$\pm$  0.00\\
 & & &10.28$\pm$ 0.05 & 30.45$\pm$  6.09 &0.20$\pm$0.05 &0.60$\pm$0.06 &0.64$\pm$0.00 &-21.50$\pm$  0.00\\
 & & &10.31$\pm$ 0.05 & 32.69$\pm$  6.54 &0.20$\pm$0.05 &0.60$\pm$0.06 &0.65$\pm$0.00 &-21.59$\pm$  0.00\\
\hline
  21739 & 1 &0.25 & 8.97$\pm$ 0.04 &  5.52$\pm$  1.10 &0.20$\pm$0.05 &0.20$\pm$0.02 &0.40$\pm$0.11 &-18.98$\pm$  0.08\\
 & & & 9.03$\pm$ 0.03 &  6.37$\pm$  1.42 &0.20$\pm$0.05 &0.20$\pm$0.02 &0.38$\pm$0.10 &-19.11$\pm$  0.07\\
 & & & 8.86$\pm$ 0.04 &  4.33$\pm$  0.87 &0.20$\pm$0.05 &0.20$\pm$0.02 &0.41$\pm$0.12 &-18.71$\pm$  0.10\\
  21739 & 2 &0.22 & 8.16$\pm$ 0.00 & 14.84$\pm$  2.97 &0.25$\pm$0.05 &0.01$\pm$0.00 &0.12$\pm$0.11 &-18.52$\pm$  0.07\\
 & & & 8.23$\pm$ 0.00 & 17.36$\pm$  3.47 &0.25$\pm$0.05 &0.01$\pm$0.00 &0.14$\pm$0.09 &-18.71$\pm$  0.07\\
 & & & 8.12$\pm$ 0.00 & 13.47$\pm$  2.69 &0.25$\pm$0.05 &0.01$\pm$0.00 &0.03$\pm$0.12 &-18.35$\pm$  0.10\\
  21739 & 3 &0.07 & 7.97$\pm$ 0.29 &  9.51$\pm$  2.38 &0.20$\pm$0.06 &0.01$\pm$0.01 &-0.04$\pm$0.13 &-18.24$\pm$  0.10\\
 & & & 8.65$\pm$ 0.07 &  1.77$\pm$  0.45 &0.05$\pm$0.05 &0.30$\pm$0.08 &0.01$\pm$0.10 &-18.49$\pm$  0.08\\
 & & & 7.87$\pm$ 0.00 &  7.53$\pm$  1.51 &0.20$\pm$0.05 &0.01$\pm$0.00 &-0.08$\pm$0.18 &-17.95$\pm$  0.14\\
  21739 & 4 &0.34 & 8.09$\pm$ 0.02 &  0.50$\pm$  0.10 &0.05$\pm$0.05 &0.30$\pm$0.03 &0.15$\pm$0.17 &-17.18$\pm$  0.10\\
 & & & 8.27$\pm$ 0.07 &  0.61$\pm$  0.12 &0.00$\pm$0.05 &0.20$\pm$0.03 &0.19$\pm$0.09 &-17.73$\pm$  0.07\\
 & & & 8.30$\pm$ 0.05 &  0.49$\pm$  0.10 &0.05$\pm$0.05 &0.50$\pm$0.06 &0.24$\pm$0.14 &-17.31$\pm$  0.10\\
  21739 & 5 &0.52 & 7.74$\pm$ 0.10 &  0.70$\pm$  0.51 &0.05$\pm$0.05 &0.09$\pm$0.03 &0.13$\pm$0.18 &-17.04$\pm$  0.13\\
 & & & 8.19$\pm$ 0.01 &  0.92$\pm$  0.18 &0.05$\pm$0.05 &0.20$\pm$0.00 &0.18$\pm$0.11 &-17.65$\pm$  0.09\\
 & & & 8.00$\pm$ 0.01 &  0.60$\pm$  0.12 &0.05$\pm$0.05 &0.20$\pm$0.00 &0.20$\pm$0.15 &-17.18$\pm$  0.12\\
  21739 & D &---- & 9.80$\pm$ 0.04 & 25.18$\pm$  5.04 &0.10$\pm$0.05 &0.30$\pm$0.05 &0.25$\pm$0.00 &-21.24$\pm$  0.00\\
 & & & 9.92$\pm$ 0.02 & 33.64$\pm$  6.73 &0.10$\pm$0.05 &0.30$\pm$0.03 &0.23$\pm$0.00 &-21.54$\pm$  0.00\\
 & & & 9.96$\pm$ 0.02 & 36.81$\pm$  7.36 &0.10$\pm$0.05 &0.30$\pm$0.03 &0.24$\pm$0.00 &-21.64$\pm$  0.00\\
\hline
  21852 & 1 &0.37 & 8.84$\pm$ 0.07 &  2.09$\pm$  0.42 &0.25$\pm$0.05 &0.40$\pm$0.07 &0.52$\pm$0.15 &-17.97$\pm$  0.12\\
 & & & 8.93$\pm$ 0.08 &  2.61$\pm$  0.52 &0.25$\pm$0.05 &0.40$\pm$0.07 &0.54$\pm$0.12 &-18.24$\pm$  0.08\\
 & & & 8.82$\pm$ 0.03 &  1.17$\pm$  0.57 &0.20$\pm$0.05 &0.70$\pm$0.03 &0.60$\pm$0.17 &-17.78$\pm$  0.14\\
  21852 & 2 &0.10 & 9.01$\pm$ 0.08 &  3.37$\pm$  0.67 &0.25$\pm$0.05 &0.20$\pm$0.03 &0.67$\pm$0.13 &-18.58$\pm$  0.09\\
 & & & 9.13$\pm$ 0.10 &  3.09$\pm$  0.62 &0.25$\pm$0.05 &0.30$\pm$0.07 &0.66$\pm$0.14 &-18.74$\pm$  0.10\\
 & & & 9.05$\pm$ 0.03 &  2.75$\pm$  0.55 &0.30$\pm$0.05 &0.50$\pm$0.03 &0.68$\pm$0.22 &-18.15$\pm$  0.15\\
  21852 & 3 &0.20 & 8.83$\pm$ 0.03 &  2.69$\pm$  0.77 &0.25$\pm$0.05 &0.30$\pm$0.02 &0.51$\pm$0.18 &-18.17$\pm$  0.12\\
 & & & 9.00$\pm$ 0.04 &  2.07$\pm$  0.41 &0.20$\pm$0.05 &0.60$\pm$0.09 &0.53$\pm$0.16 &-18.40$\pm$  0.10\\
 & & & 8.54$\pm$ 0.03 &  2.07$\pm$  0.41 &0.25$\pm$0.05 &0.20$\pm$0.02 &0.46$\pm$0.29 &-17.68$\pm$  0.22\\
  21852 & 4 &0.09 & 9.34$\pm$ 0.03 &  6.67$\pm$  1.33 &0.40$\pm$0.05 &0.40$\pm$0.03 &0.80$\pm$0.14 &-18.62$\pm$  0.09\\
 & & & 9.40$\pm$ 0.03 &  5.22$\pm$  1.04 &0.35$\pm$0.05 &0.60$\pm$0.06 &0.77$\pm$0.14 &-18.78$\pm$  0.09\\
 & & & 9.27$\pm$ 0.04 & 10.98$\pm$  2.20 &0.65$\pm$0.05 &0.20$\pm$0.02 &1.05$\pm$0.37 &-17.80$\pm$  0.25\\
  21852 & D &---- & 9.99$\pm$ 0.15 & 10.83$\pm$  2.91 &0.15$\pm$0.06 &0.30$\pm$0.07 &0.60$\pm$0.00 &-21.12$\pm$  0.00\\
 & & &10.35$\pm$ 0.04 & 40.35$\pm$  8.07 &0.25$\pm$0.05 &0.70$\pm$0.07 &0.61$\pm$0.00 &-21.42$\pm$  0.00\\
 & & &10.41$\pm$ 0.03 & 46.11$\pm$  9.22 &0.25$\pm$0.05 &0.70$\pm$0.06 &0.61$\pm$0.00 &-21.55$\pm$  0.00\\
\hline
  22284 & 1 &0.24 & 8.86$\pm$ 0.06 &  2.92$\pm$  1.79 &0.10$\pm$0.05 &0.30$\pm$0.05 &0.26$\pm$0.14 &-18.91$\pm$  0.10\\
 & & & 8.96$\pm$ 0.03 &  5.47$\pm$  1.22 &0.15$\pm$0.05 &0.20$\pm$0.02 &0.30$\pm$0.10 &-19.16$\pm$  0.08\\
 & & & 8.72$\pm$ 0.04 &  2.08$\pm$  0.42 &0.10$\pm$0.05 &0.30$\pm$0.05 &0.33$\pm$0.19 &-18.57$\pm$  0.14\\
  22284 & 2 &0.11 & 9.47$\pm$ 0.11 &  3.76$\pm$  0.75 &0.25$\pm$0.05 &0.40$\pm$0.09 &0.84$\pm$0.14 &-19.28$\pm$  0.10\\
 & & & 9.42$\pm$ 0.18 &  2.91$\pm$  0.58 &0.20$\pm$0.05 &0.30$\pm$0.11 &0.78$\pm$0.10 &-19.47$\pm$  0.08\\
 & & & 9.38$\pm$ 0.12 &  1.89$\pm$  0.38 &0.25$\pm$0.05 &0.60$\pm$0.13 &0.92$\pm$0.17 &-18.84$\pm$  0.13\\
  22284 & 3 &0.39 & 7.70$\pm$ 0.00 &  5.11$\pm$  1.02 &0.20$\pm$0.05 &0.01$\pm$0.00 &-0.16$\pm$0.20 &-17.43$\pm$  0.15\\
 & & & 8.43$\pm$ 0.07 &  3.07$\pm$  0.62 &0.15$\pm$0.05 &0.10$\pm$0.02 &0.12$\pm$0.12 &-18.21$\pm$  0.08\\
 & & & 7.61$\pm$ 0.00 &  4.13$\pm$  0.83 &0.20$\pm$0.05 &0.01$\pm$0.00 &0.00$\pm$0.26 &-17.32$\pm$  0.20\\
  22284 & D &---- &10.42$\pm$ 0.01 & 64.88$\pm$ 12.98 &0.20$\pm$0.05 &0.50$\pm$0.02 &0.48$\pm$0.00 &-21.99$\pm$  0.00\\
 & & &10.51$\pm$ 0.12 & 51.32$\pm$ 10.26 &0.15$\pm$0.05 &0.80$\pm$0.21 &0.48$\pm$0.00 &-22.15$\pm$  0.00\\
 & & &10.36$\pm$ 0.11 & 51.99$\pm$ 10.40 &0.15$\pm$0.05 &0.30$\pm$0.07 &0.48$\pm$0.00 &-22.23$\pm$  0.00\\
\hline
  23013 & 1 &0.04 &10.03$\pm$ 0.01 & 63.07$\pm$ 12.61 &0.55$\pm$0.05 &0.20$\pm$0.00 &0.84$\pm$0.12 &-20.07$\pm$  0.09\\
 & & &10.08$\pm$ 0.04 & 48.06$\pm$  9.61 &0.50$\pm$0.05 &0.30$\pm$0.03 &0.81$\pm$0.11 &-20.19$\pm$  0.07\\
 & & & 9.89$\pm$ 0.04 & 30.98$\pm$  6.20 &0.50$\pm$0.05 &0.30$\pm$0.05 &0.84$\pm$0.16 &-19.76$\pm$  0.11\\
  23013 & 2 &0.35 & 9.01$\pm$ 0.07 & 11.72$\pm$  2.38 &0.30$\pm$0.05 &0.10$\pm$0.02 &0.42$\pm$0.15 &-18.97$\pm$  0.10\\
 & & & 9.25$\pm$ 0.05 & 17.40$\pm$  3.70 &0.30$\pm$0.05 &0.20$\pm$0.05 &0.46$\pm$0.11 &-19.28$\pm$  0.09\\
 & & & 8.80$\pm$ 0.02 &  9.06$\pm$  1.81 &0.30$\pm$0.05 &0.08$\pm$0.00 &0.39$\pm$0.18 &-18.65$\pm$  0.14\\
  23013 & 3 &0.19 & 9.29$\pm$ 0.01 & 22.52$\pm$  4.50 &0.45$\pm$0.05 &0.10$\pm$0.00 &0.66$\pm$0.17 &-19.11$\pm$  0.11\\
 & & & 9.49$\pm$ 0.01 & 18.45$\pm$  3.69 &0.40$\pm$0.05 &0.20$\pm$0.00 &0.65$\pm$0.13 &-19.39$\pm$  0.10\\
 & & & 9.12$\pm$ 0.06 & 16.62$\pm$  7.61 &0.45$\pm$0.05 &0.09$\pm$0.02 &0.63$\pm$0.24 &-18.71$\pm$  0.17\\
  23013 & 4 &0.65 & 7.02$\pm$ 0.00 &  1.06$\pm$  0.21 &0.00$\pm$0.05 &0.01$\pm$0.00 &-0.71$\pm$0.31 &-16.34$\pm$  0.29\\
 & & & 8.25$\pm$ 0.18 &  2.04$\pm$  8.36 &0.10$\pm$0.05 &0.10$\pm$0.07 &0.17$\pm$0.10 &-18.03$\pm$  0.06\\
 & & & 8.03$\pm$ 0.02 &  1.74$\pm$  0.35 &0.10$\pm$0.05 &0.07$\pm$0.00 &0.11$\pm$0.13 &-17.67$\pm$  0.09\\
  23013 & D &---- &10.63$\pm$ 0.10 &131.35$\pm$ 26.27 &0.30$\pm$0.05 &0.40$\pm$0.08 &0.61$\pm$0.00 &-22.29$\pm$  0.00\\
 & & &10.73$\pm$ 0.10 &162.39$\pm$ 32.48 &0.30$\pm$0.05 &0.40$\pm$0.08 &0.62$\pm$0.00 &-22.51$\pm$  0.00\\
 & & &10.76$\pm$ 0.03 &176.46$\pm$ 35.29 &0.30$\pm$0.05 &0.40$\pm$0.02 &0.62$\pm$0.00 &-22.61$\pm$  0.00\\
\hline
  24033 & 1 &0.30 & 9.09$\pm$ 0.04 & 49.43$\pm$ 31.14 &0.25$\pm$0.05 &0.40$\pm$0.03 &0.50$\pm$0.14 &-18.88$\pm$  0.11\\
 & & & 9.27$\pm$ 0.10 & 54.11$\pm$ 27.27 &0.25$\pm$0.05 &0.60$\pm$0.17 &0.54$\pm$0.12 &-19.10$\pm$  0.09\\
 & & & 8.87$\pm$ 0.04 & 14.83$\pm$  5.16 &0.25$\pm$0.05 &0.20$\pm$0.02 &0.43$\pm$0.16 &-18.65$\pm$  0.12\\
  24033 & 2 &0.12 & 9.50$\pm$ 0.10 &  9.66$\pm$  3.97 &0.35$\pm$0.05 &0.40$\pm$0.11 &0.75$\pm$0.16 &-19.22$\pm$  0.10\\
 & & & 9.56$\pm$ 0.15 & 11.05$\pm$  7.42 &0.35$\pm$0.05 &0.40$\pm$0.16 &0.75$\pm$0.15 &-19.38$\pm$  0.09\\
 & & & 9.20$\pm$ 0.03 &  6.37$\pm$  1.27 &0.35$\pm$0.05 &0.30$\pm$0.02 &0.71$\pm$0.21 &-18.72$\pm$  0.17\\
  24033 & 3 &0.07 &10.23$\pm$ 0.09 & 41.96$\pm$  8.39 &0.65$\pm$0.05 &0.50$\pm$0.06 &1.19$\pm$0.14 &-19.63$\pm$  0.09\\
 & & &10.26$\pm$ 0.09 & 25.24$\pm$  5.05 &0.55$\pm$0.05 &0.90$\pm$0.18 &1.14$\pm$0.11 &-19.75$\pm$  0.08\\
 & & &10.11$\pm$ 0.10 & 14.20$\pm$  5.08 &0.70$\pm$0.09 &0.30$\pm$0.02 &1.45$\pm$0.26 &-19.07$\pm$  0.15\\
  24033 & 4 &0.27 & 9.45$\pm$ 0.04 & 11.23$\pm$  2.25 &0.50$\pm$0.05 &0.30$\pm$0.03 &0.70$\pm$0.17 &-18.58$\pm$  0.11\\
 & & & 9.55$\pm$ 0.08 & 10.77$\pm$  2.15 &0.45$\pm$0.05 &0.40$\pm$0.07 &0.71$\pm$0.14 &-18.86$\pm$  0.09\\
 & & & 9.20$\pm$ 0.00 &  9.48$\pm$  1.90 &0.50$\pm$0.05 &0.20$\pm$0.00 &0.73$\pm$0.23 &-18.23$\pm$  0.15\\
  24033 & D &---- &10.70$\pm$ 0.16 & 78.01$\pm$ 20.17 &0.35$\pm$0.06 &0.80$\pm$0.20 &0.75$\pm$0.00 &-21.76$\pm$  0.00\\
 & & &10.79$\pm$ 0.15 &111.00$\pm$ 31.07 &0.35$\pm$0.06 &0.70$\pm$0.16 &0.77$\pm$0.00 &-22.10$\pm$  0.00\\
 & & &10.85$\pm$ 0.15 &125.39$\pm$ 33.86 &0.35$\pm$0.06 &0.70$\pm$0.16 &0.78$\pm$0.00 &-22.23$\pm$  0.00\\
\hline
  24684 & 1 &0.20 & 8.86$\pm$ 0.14 &  2.88$\pm$  3.10 &0.35$\pm$0.05 &0.30$\pm$0.12 &0.74$\pm$0.12 &-17.85$\pm$  0.09\\
 & & & 9.20$\pm$ 0.13 &  3.30$\pm$  3.30 &0.35$\pm$0.05 &0.60$\pm$0.24 &0.92$\pm$0.10 &-18.33$\pm$  0.07\\
 & & & 8.70$\pm$ 0.06 &  2.94$\pm$  0.59 &0.35$\pm$0.05 &0.20$\pm$0.03 &0.63$\pm$0.16 &-17.61$\pm$  0.12\\
  24684 & 2 &0.04 &10.48$\pm$ 0.31 &  0.00$\pm$  0.00 &0.30$\pm$0.40 &3.00$\pm$0.00 &2.47$\pm$0.13 &-19.38$\pm$  0.06\\
 & & &10.87$\pm$ 0.12 &  3.27$\pm$  2.42 &0.60$\pm$0.20 &3.00$\pm$0.00 &2.26$\pm$0.10 &-19.51$\pm$  0.07\\
 & & &10.93$\pm$ 0.01 &  3.78$\pm$  2.33 &0.75$\pm$0.14 &3.00$\pm$0.00 &2.60$\pm$0.16 &-19.10$\pm$  0.07\\
  24684 & 3 &0.32 & 7.75$\pm$ 0.03 &  0.91$\pm$  0.18 &0.10$\pm$0.05 &0.07$\pm$0.01 &0.03$\pm$0.18 &-16.91$\pm$  0.15\\
 & & & 8.76$\pm$ 0.04 &  0.53$\pm$  0.11 &0.05$\pm$0.05 &1.40$\pm$0.19 &0.63$\pm$0.08 &-17.87$\pm$  0.06\\
 & & & 8.05$\pm$ 0.04 &  0.45$\pm$  0.09 &0.05$\pm$0.05 &0.30$\pm$0.02 &0.37$\pm$0.17 &-17.21$\pm$  0.12\\
  24684 & 4 &0.13 & 9.37$\pm$ 0.07 &  1.69$\pm$  0.38 &0.40$\pm$0.05 &1.80$\pm$0.23 &1.22$\pm$0.17 &-17.83$\pm$  0.11\\
 & & & 9.74$\pm$ 0.01 &  2.43$\pm$  0.49 &0.40$\pm$0.05 &3.00$\pm$0.05 &1.27$\pm$0.10 &-18.31$\pm$  0.08\\
 & & & 9.37$\pm$ 0.01 &  1.19$\pm$  0.24 &0.35$\pm$0.05 &2.60$\pm$0.09 &1.17$\pm$0.19 &-17.74$\pm$  0.13\\
  24684 & 5 &0.31 & 8.63$\pm$ 0.02 &  1.70$\pm$  0.34 &0.35$\pm$0.05 &0.30$\pm$0.03 &0.71$\pm$0.18 &-17.28$\pm$  0.13\\
 & & & 9.20$\pm$ 0.04 &  2.23$\pm$  0.61 &0.35$\pm$0.05 &0.90$\pm$0.09 &0.97$\pm$0.11 &-18.00$\pm$  0.06\\
 & & & 8.72$\pm$ 0.04 &  0.94$\pm$  0.49 &0.30$\pm$0.05 &0.70$\pm$0.14 &0.83$\pm$0.24 &-17.15$\pm$  0.15\\
  24684 & D &---- &11.01$\pm$ 0.09 & 12.58$\pm$  5.73 &0.25$\pm$0.12 &2.20$\pm$0.17 &1.36$\pm$0.01 &-21.71$\pm$  0.00\\
 & & &11.06$\pm$ 0.07 & 10.93$\pm$  6.60 &0.20$\pm$0.13 &2.40$\pm$0.13 &1.36$\pm$0.00 &-21.92$\pm$  0.00\\
 & & &11.10$\pm$ 0.09 & 11.86$\pm$  6.59 &0.20$\pm$0.13 &2.40$\pm$0.13 &1.37$\pm$0.00 &-22.01$\pm$  0.00\\
\hline
  24919 & 1 &0.22 & 9.23$\pm$ 0.01 & 21.62$\pm$  4.32 &0.30$\pm$0.05 &0.09$\pm$0.00 &0.30$\pm$0.18 &-19.60$\pm$  0.12\\
 & & & 9.30$\pm$ 0.10 & 37.34$\pm$  7.47 &0.35$\pm$0.05 &0.06$\pm$0.02 &0.39$\pm$0.15 &-19.88$\pm$  0.12\\
 & & & 9.21$\pm$ 0.00 & 18.85$\pm$  3.77 &0.30$\pm$0.05 &0.10$\pm$0.00 &0.35$\pm$0.19 &-19.50$\pm$  0.16\\
  24919 & 2 &0.07 & 9.99$\pm$ 0.03 & 29.81$\pm$  5.96 &0.40$\pm$0.05 &0.40$\pm$0.03 &0.93$\pm$0.15 &-20.26$\pm$  0.12\\
 & & &10.06$\pm$ 0.03 & 35.04$\pm$  7.01 &0.40$\pm$0.05 &0.40$\pm$0.03 &0.90$\pm$0.12 &-20.42$\pm$  0.10\\
 & & & 9.76$\pm$ 0.00 & 23.24$\pm$  4.65 &0.40$\pm$0.05 &0.30$\pm$0.00 &0.91$\pm$0.20 &-19.92$\pm$  0.16\\
  24919 & 3 &0.25 & 9.30$\pm$ 0.08 & 37.13$\pm$  7.43 &0.55$\pm$0.05 &0.06$\pm$0.02 &0.76$\pm$0.18 &-19.03$\pm$  0.15\\
 & & & 9.60$\pm$ 0.04 & 23.63$\pm$  4.73 &0.45$\pm$0.05 &0.20$\pm$0.02 &0.75$\pm$0.13 &-19.47$\pm$  0.11\\
 & & & 9.26$\pm$ 0.02 & 25.65$\pm$  5.13 &0.50$\pm$0.05 &0.08$\pm$0.00 &0.70$\pm$0.23 &-18.91$\pm$  0.17\\
  24919 & 4 &0.25 & 9.69$\pm$ 0.00 &494.21$\pm$ 98.84 &0.95$\pm$0.05 &0.01$\pm$0.00 &1.07$\pm$0.16 &-19.26$\pm$  0.14\\
 & & &10.02$\pm$ 0.02 & 42.12$\pm$  8.42 &0.60$\pm$0.05 &0.30$\pm$0.03 &0.97$\pm$0.13 &-19.63$\pm$  0.09\\
 & & & 9.72$\pm$ 0.00 &530.72$\pm$106.14 &1.00$\pm$0.05 &0.01$\pm$0.00 &1.04$\pm$0.23 &-19.06$\pm$  0.19\\
  24919 & 5 &0.33 & 9.27$\pm$ 0.04 &  7.46$\pm$  1.49 &0.35$\pm$0.05 &0.30$\pm$0.03 &0.56$\pm$0.20 &-18.76$\pm$  0.17\\
 & & & 9.45$\pm$ 0.04 & 11.22$\pm$  2.24 &0.35$\pm$0.05 &0.30$\pm$0.05 &0.63$\pm$0.14 &-19.30$\pm$  0.12\\
 & & & 9.09$\pm$ 0.00 &  7.22$\pm$  1.44 &0.35$\pm$0.05 &0.20$\pm$0.00 &0.60$\pm$0.30 &-18.57$\pm$  0.25\\
  24919 & D &---- &10.92$\pm$ 0.05 &334.27$\pm$ 66.85 &0.40$\pm$0.05 &0.30$\pm$0.05 &0.74$\pm$0.00 &-22.80$\pm$  0.00\\
 & & &11.02$\pm$ 0.02 &423.67$\pm$ 84.73 &0.40$\pm$0.05 &0.30$\pm$0.03 &0.74$\pm$0.00 &-23.03$\pm$  0.00\\
 & & &11.06$\pm$ 0.03 &463.33$\pm$ 92.67 &0.40$\pm$0.05 &0.30$\pm$0.02 &0.74$\pm$0.00 &-23.13$\pm$  0.00\\
\hline
  26067 & 1 &0.09 & 9.35$\pm$ 0.07 & 23.57$\pm$  7.39 &0.20$\pm$0.05 &0.60$\pm$0.10 &0.55$\pm$0.14 &-19.44$\pm$  0.11\\
 & & & 9.36$\pm$ 0.04 &  9.27$\pm$  1.85 &0.20$\pm$0.05 &0.40$\pm$0.03 &0.56$\pm$0.14 &-19.57$\pm$  0.10\\
 & & & 9.14$\pm$ 0.02 &  5.56$\pm$  1.21 &0.20$\pm$0.05 &0.30$\pm$0.02 &0.52$\pm$0.15 &-19.18$\pm$  0.15\\
  26067 & 2 &0.13 & 9.87$\pm$ 0.06 & 13.36$\pm$  2.67 &0.40$\pm$0.05 &0.70$\pm$0.10 &0.83$\pm$0.12 &-19.53$\pm$  0.09\\
 & & & 9.93$\pm$ 0.08 & 10.71$\pm$  2.14 &0.35$\pm$0.05 &1.00$\pm$0.20 &0.81$\pm$0.12 &-19.65$\pm$  0.10\\
 & & & 9.77$\pm$ 0.06 & 10.50$\pm$  2.10 &0.40$\pm$0.05 &0.70$\pm$0.10 &0.85$\pm$0.20 &-19.28$\pm$  0.15\\
  26067 & D &---- &10.55$\pm$ 0.05 & 49.82$\pm$  9.96 &0.25$\pm$0.05 &0.90$\pm$0.09 &0.66$\pm$0.00 &-21.72$\pm$  0.00\\
 & & &10.61$\pm$ 0.11 & 64.39$\pm$ 12.88 &0.25$\pm$0.05 &0.80$\pm$0.20 &0.67$\pm$0.00 &-21.97$\pm$  0.00\\
 & & &10.64$\pm$ 0.11 & 68.62$\pm$ 13.72 &0.25$\pm$0.05 &0.80$\pm$0.20 &0.67$\pm$0.00 &-22.04$\pm$  0.00\\
\hline
  27101 & 1 &0.06 & 9.61$\pm$ 0.05 &  2.91$\pm$  0.58 &0.30$\pm$0.05 &1.80$\pm$0.20 &1.02$\pm$0.10 &-18.81$\pm$  0.07\\
 & & & 9.63$\pm$ 0.04 &  3.91$\pm$  0.78 &0.30$\pm$0.05 &1.40$\pm$0.16 &0.98$\pm$0.10 &-18.95$\pm$  0.06\\
 & & & 9.39$\pm$ 0.07 &  2.58$\pm$  2.85 &0.30$\pm$0.05 &1.20$\pm$0.28 &0.99$\pm$0.14 &-18.55$\pm$  0.10\\
  27101 & 2 &0.17 & 8.52$\pm$ 0.01 &  5.32$\pm$  1.06 &0.30$\pm$0.05 &0.07$\pm$0.00 &0.50$\pm$0.13 &-18.08$\pm$  0.09\\
 & & & 8.71$\pm$ 0.64 &  5.89$\pm$ 44.74 &0.30$\pm$0.05 &0.10$\pm$0.09 &0.54$\pm$0.09 &-18.35$\pm$  0.06\\
 & & & 8.58$\pm$ 0.02 &  4.32$\pm$  0.86 &0.30$\pm$0.05 &0.10$\pm$0.01 &0.55$\pm$0.14 &-18.01$\pm$  0.10\\
  27101 & 3 &0.17 & 8.06$\pm$ 0.00 & 11.78$\pm$  2.36 &0.40$\pm$0.05 &0.01$\pm$0.00 &0.10$\pm$0.12 &-17.45$\pm$  0.09\\
 & & & 8.18$\pm$ 0.00 & 15.62$\pm$  3.12 &0.40$\pm$0.05 &0.01$\pm$0.00 &0.26$\pm$0.09 &-17.89$\pm$  0.06\\
 & & & 7.98$\pm$ 0.00 &  9.83$\pm$  1.97 &0.40$\pm$0.05 &0.01$\pm$0.00 &0.10$\pm$0.17 &-17.25$\pm$  0.13\\
  27101 & 4 &0.17 & 9.00$\pm$ 0.07 &  0.82$\pm$  0.21 &0.20$\pm$0.05 &1.60$\pm$0.28 &0.81$\pm$0.16 &-17.83$\pm$  0.10\\
 & & & 9.14$\pm$ 0.04 &  1.26$\pm$  1.11 &0.20$\pm$0.05 &1.40$\pm$0.27 &0.78$\pm$0.08 &-18.16$\pm$  0.06\\
 & & & 8.77$\pm$ 0.16 &  1.21$\pm$  1.16 &0.25$\pm$0.05 &0.60$\pm$0.31 &0.77$\pm$0.17 &-17.68$\pm$  0.13\\
  27101 & 5 &0.27 & 8.59$\pm$ 0.07 &  0.41$\pm$  0.46 &0.10$\pm$0.05 &1.20$\pm$0.28 &0.61$\pm$0.15 &-17.36$\pm$  0.10\\
 & & & 8.80$\pm$ 0.06 &  0.89$\pm$  0.18 &0.15$\pm$0.05 &0.90$\pm$0.12 &0.64$\pm$0.10 &-17.82$\pm$  0.07\\
 & & & 8.16$\pm$ 0.16 &  0.58$\pm$  1.30 &0.15$\pm$0.05 &0.30$\pm$0.18 &0.45$\pm$0.21 &-16.95$\pm$  0.15\\
  27101 & D &---- &10.10$\pm$ 0.06 & 50.42$\pm$ 15.87 &0.30$\pm$0.05 &0.30$\pm$0.05 &0.70$\pm$0.01 &-21.19$\pm$  0.00\\
 & & &10.21$\pm$ 0.05 & 64.68$\pm$ 17.49 &0.30$\pm$0.05 &0.30$\pm$0.04 &0.69$\pm$0.00 &-21.46$\pm$  0.00\\
 & & &10.25$\pm$ 0.03 & 70.69$\pm$ 14.14 &0.30$\pm$0.05 &0.30$\pm$0.02 &0.70$\pm$0.00 &-21.56$\pm$  0.00\\
\hline
\end{longtable}
\footnotetext[1]{Numbers stand for ID of clumps in the z-band images of Figure \ref{fig:mosaic}, while ``D'' stands for the diffuse ``disk'' component.}
\footnotetext[2]{For each clump, properties derived under different background 
subtraction schemes are listed: global constant background (1st line), zero background (2nd line) and local background (3rd line).}

\begin{table*}[htbp]
\begin{minipage}[center]{\textwidth}
\caption{Comparison o9 Derived Properties of Three Galaxies in Our Sample and Elmegreen \& Elmegreen (2005) \label{tb:comp}} \footnote[0]{Note: For each galaxy, the first line shows the data in our work, while the second line shows the data in Elmegreen et al. (2005). In this table, we have applied the relations in Sec. \ref{disc:other} to convert the Chabrier IMF to the Salpeter IMF.}
\begin{tabular}{cccccccccc}
\hline\hline
Galaxy & ID & z & ${\rm N_{clump}}$ & Age$_{\rm clump}$ & Age$_{\rm disk}$ & $<{\rm M_{clump}}>$ & ${\rm f_{mass}}$  & ${\rm M_{gal}}$ & $<{\rm SFR_{clump}}>$  \\
&  &  &  & (Gyr) & (Gyr) & ${\rm 10^9 M_\odot}$ & & ${\rm 10^{10} M_\odot}$ & ${\rm M_\odot yr^{-1}}$ \\
\hline
1 &  21739 &  1.765 & 5  & 0.12 & 0.30 & 0.27 & 0.16 & 0.8 & 6.2 \\
  &  3465+ &  2.4   & 11 & 0.22 & 1.80 & 0.87 & 0.28 & 3.5 & 4.5 \\
\hline
2 &  22284 &  1.767 & 3  & 0.24 & 0.50 & 1.25 & 0.12 & 3.2 & 3.9 \\
  &  3483  &  2.2   & 12 & 0.26 & 2.02 & 1.31 & 0.32 & 4.9 & 1.0 \\
\hline
3 &  27101 &  1.570 & 5  & 0.94 & 0.30 & 1.18 & 0.36 & 1.6 & 4.3 \\
  &  6462+ &  2.8   & 8  & 0.31 & 2.82 & 1.57 & 0.22 & 5.7 & 0.9 \\
\hline
\hline
\end{tabular}
\end{minipage}
\end{table*}

\end{document}